\documentclass[11pt,english]{amsart}
\usepackage{ae,aecompl}
\usepackage[T1]{fontenc}
\usepackage[latin9]{inputenc}
\usepackage{geometry}
\usepackage{array}
\usepackage{float}
\usepackage{multirow}
\usepackage{amstext}
\usepackage{amsthm}
\usepackage{amssymb}
\usepackage{graphicx}
\usepackage{setspace}
\usepackage{esint}
\usepackage{graphicx,amsmath,amssymb,epsfig}
\usepackage{amsmath}
\usepackage{amsfonts}
\usepackage{geometry}
\usepackage{setspace}
\usepackage{lscape}
\usepackage{caption}
\usepackage{rotating}
\usepackage{xcolor}
\usepackage{subfig}
\usepackage{babel}
\usepackage[authoryear]{natbib}
\usepackage{mathrsfs}
\usepackage{bm}

\makeatletter
\floatstyle{ruled}
\numberwithin{equation}{section}
\numberwithin{figure}{section}

\theoremstyle{plain}
\newtheorem{theorem}{Theorem}

\newtheorem{condition}{Condition}

\newtheorem{lemma}{Lemma}

\numberwithin{equation}{section}
\theoremstyle{definition}

\@ifundefined{showcaptionsetup}{}{ \PassOptionsToPackage{caption=false}{subfig}}
\makeatother
\providecommand{\remarkname}{Remark}

\begin{document}

\title{Dual regression}

\author{Richard H. Spady$^\dag$ \and Sami Stouli$^\S$}

%\date{\today }

\thanks{First ArXiv version: 25 October 2012. 
We are indebted to Andrew Chesher for many fruitful
discussions %on the topic of this paper, 
and to Roger Koenker for encouragement
at a key early stage. We also thank Dennis Kristensen, Lars Nesheim, David Pacini, 
Jelmer Ypma, Yanos Zylberberg, the editor, the associate editor, two anonymous referees 
and participants to seminars and conferences for helpful comments that 
considerably improved the paper. Sami Stouli gratefully acknowledges the financial
support of the UK Economic and Social Research Council and of the
Royal Economic Society.}
%\smallskip
\thanks{\noindent $^\dag$Nuffield College, Oxford, and Department of Economics, Johns Hopkins University, richard.spady@nuffield.ox.ac.uk}
\thanks{ $\S$ Department of Economics, University of Bristol, s.stouli@bristol.ac.uk}

\begin{abstract}
We propose dual regression as an alternative to the quantile regression
process for the global estimation of conditional distribution functions
under minimal assumptions. Dual regression provides all the interpretational
power of the quantile regression process while avoiding the need for
repairing the intersecting conditional quantile surfaces that quantile
regression often produces in practice. Our approach introduces a mathematical
programming characterization of conditional distribution functions
which, in its simplest form, is the dual program of a simultaneous
estimator for linear location-scale models. We apply our general characterization
to the specification and estimation of a flexible class of conditional
distribution functions, and present asymptotic theory for the corresponding
empirical dual regression process.
\end{abstract}
\maketitle
\noindent \textsc{keywords:} Conditional distribution; Duality; Monotonicity; Quantile regression; Method of moments; Mathematical programming; Convex approximation.

\section{Introduction}

Let $Y$ be a continuously distributed random variable and $X$ a random
vector. Then the conditional distribution function of $Y$ given $X$,
written $U=F_{Y\mid X}(Y\mid X)$, has three properties: %(i) 
$U$ is standard uniform, %(ii) 
$U$ is independent of $X$, and %(iii)
 $F_{Y\mid X}(y\mid x)$
is strictly increasing in $y$ for any value $x$ of $X$. We will
refer to these three properties as uniformity, independence
and monotonicity. For some specified mean zero and unit variance
distribution function $F$ with support the real line and inverse
function  $F^{-1}$, define $\varepsilon=F^{-1}\{F_{Y\mid X}(Y\mid X)\}$.
Then $\varepsilon$ satisfies independence and monotonicity, has distribution
$F$, and is transformed to uniformity by taking $U=F(\varepsilon)$.

\begin{sloppy}
If we have a sample of $n$ points $\{(x_{i},y_{i})\}_{i=1}^{n}$
drawn from the joint distribution $F_{YX}(Y,X)$, how might we estimate
the $n$ values $\varepsilon_{i}=F^{-1}\{F_{Y\mid X}(y_{i}\mid x_{i})\}$
using only the requirement that the estimate displays independence
and monotonicity, and has distribution $F$? We explore this question
by formulating a sequence of mathematical programming problems that
embodies these requirements, with each element of this sequence providing
an asymptotically valid characterization of an increasingly flexible
class of conditional distribution functions.
\par\end{sloppy}

The use of dual is thus motivated by the general observation that
the estimation problem for a conditional distribution function $F_{Y\mid X}$
indexed by a parameter $\theta$ is usually formulated in terms of
a procedure that obtains $\theta$ directly and $F_{Y\mid X}$ as a byproduct
that follows from a calculation from the representation evaluated
at a specific value of $\theta$. A classical example is the linear
location shift model $F_{Y\mid X}(y_{i}\mid x_{i})=F\{(y_{i}-\beta^\textrm{T} x_{i})/\sigma\}$,
for which the parameter vector  $\theta=(\beta,\sigma)^{\textrm{T}}$
needs to be estimated in order to obtain the $n$ values $\varepsilon_{i}=(y_{i}-\beta^\textrm{T} x_{i})/\sigma$.
Here we turn that process around, obtaining $\varepsilon_{i}$ first
(from a mathematical programming problem) and backing out $\theta$
afterwards, if at all. 

In its simplest form, dual regression augments the median
regression dual programming problem \citep{KB:1978} with global
second moment orthogonality constraints, while expanding the support
of parameter values from the unit interval to the real line. Adding
further global orthogonality constraints gives rise to a sequence
of augmented, generalized dual regression programs. Although each
of these programs seeks only to find the $n$ values $\varepsilon_{i}=F^{-1}(u_{i})$,
their first-order conditions show that the assignment of these $n$
values corresponds to a sequence of augmented location-scale representations,
the simplest element of which is a linear heteroscedastic model. Moreover,
their second-order conditions are equivalent to monotonicity, so 
%that
optimal dual regression solutions are free of quantile-crossing
problems.
% that the quantile regression process sometimes encounters
%in practice. 

To each element of the sequence of dual programs corresponds a convex
primal problem, both nontrivial to determine and difficult to implement,
the convexity of which guarantees uniqueness of optimal dual regression
solutions. For a given specification of $F_{Y\mid X}(Y\mid X)$, the first-order
conditions of the corresponding primal problem also describe necessary
and sufficient conditions for independence of the associated dual
solutions. Thus our dual formulation reveals a sequence of convex
optimization problems, gives a feasible and direct implementation of each
of them, and uniquely characterizes the family of associated globally
monotone representations, which can then be used as complete estimates
of a flexible class of conditional distribution functions.

\section{Basics}\label{sec:Basics}

\subsection{The dual regression problem}

We introduce the basic principles underlying our general
method by first providing a new characterization of the conditional
distribution function $F_{Y\mid X}(Y\mid X)$ associated with the linear
location-scale model
\begin{equation}
Y=\beta_{1}^\textrm{T} X+(\beta_{2}^\textrm{T} X)\varepsilon,\quad\beta_{2}^\textrm{T} X>0,\quad\varepsilon \mid X\sim F,\label{eq:DGP loc-scale}
\end{equation}
where $X$ is a $K\times1$ vector of explanatory variables including
an intercept, and $F$ a mean zero and unit variance cumulative distribution
function over the real line.

Suppose that we observe a sample of $n$ identically and independently
distributed realizations $\{(y_{i},x_{i})\}_{i=1}^{n}$ generated
according to model (\ref{eq:DGP loc-scale}). The primary population
target of our analysis is
\[
\varepsilon_{i}=\frac{y_{i}-\beta_{1}^\textrm{T} x_{i}}{\beta_{2}^\textrm{T} x_{i}}=F^{-1}\{F_{Y\mid X}(y_{i}\mid x_{i})\} \quad(i=1,\ldots,n),
\]
knowledge of which is equivalent to knowledge of the $n$ values $F_{Y\mid X}(y_{i}\mid x_{i})$
up to the monotone transformation $F$.

Let $\lambda=(\lambda_{1},\lambda_{2})^{\textrm{T}}\in\mathbb{R}^{2\times K}$
and $e_{o}\in\mathbb{R}^{n}$ satisfy the system of $n$ equations
and $n$ inequality constraints
\begin{equation}
y_{i}=\lambda_{1}^\textrm{T} x_{i}+(\lambda_{2}^\textrm{T} x_{i})e_{oi},\quad\lambda_{2}^\textrm{T} x_{i}>0 \quad(i=1,\ldots,n),\label{eq:loc-scale mod}
\end{equation}
where $e_{o}$ further satisfies the $2\times K$ orthogonality conditions
$\sum_{i=1}^{n}x_{i}e_{oi}=0$ and $\sum_{i=1}^{n}x_{i}(e_{oi}^{2}-1)=0$.
Since $x_{i}$ includes an intercept, the sample moments of $e_{o}$
and $e_{o}^{2}$ are $0$ and $1$, and $e_{o}$ and $e_{o}^{2}$
are orthogonal to each column of the $n\times K$ matrix $(x_{1},\ldots,x_{n})^{\textrm{T}}$
of explanatory variables. %The primary sample targets of our analysis
%are $e_{o}$ and its empirical distribution function, and our goal is thus to provide
We propose a characterization of
the sequence of vectors $e_{o}$ that satisfy representation
(\ref{eq:loc-scale mod}) and the associated orthogonality constraints
for each $n$. The %associated 
corresponding sequence of empirical distribution functions then provides an asymptotically valid
characterization of the conditional distribution function $F_{Y\mid X}(Y\mid X)$ corresponding to the data-generating process (\ref{eq:DGP loc-scale}). 
As a by-product of this approach, we simultaneously obtain a characterization of the parameter vector $\lambda$ in (\ref{eq:loc-scale mod}),  
which then provides a consistent estimator of the population parameter $\beta$ in (\ref{eq:DGP loc-scale}).

For each $x_{i}$, with the scale function $\lambda_{2}^\textrm{T} x_{i} > 0$,
%being strictly positive, 
$y_{i}$ is an increasing function of $e_{oi}$,
and to representation (\ref{eq:loc-scale mod}) corresponds a convex
function %$C(x_{i},e_{oi},\lambda)$ defined as 
\[
C(x_{i},e_{oi},\lambda)=\int_{0}^{e_{oi}}\{\lambda_{1}^\textrm{T} x_{i}+(\lambda_{2}^\textrm{T} x_{i})s\}ds=(\lambda_{1}^\textrm{T} x_{i})e_{oi}+\frac{1}{2}(\lambda_{2}^\textrm{T} x_{i})e_{oi}^{2} \quad(e_{oi}\in\mathbb{R}),
\]
and whose quadratic form corresponds to a location-scale representation
for $F_{Y\mid X}(Y\mid X)$. Letting $y$ be the $n\times1$ vector
of dependent variable values, and assuming knowledge of $\lambda$
and $e_{o}$, we consider assigning a value $e_{i}$ to each observation
in the sample by maximizing the correlation between $y$ and $e=(e_{1},\ldots,e_{n})^{\textrm{T}}$
subject to a constraint that embodies the properties of $e_{o}$:
\begin{equation}
\max_{e\in\mathbb{R}^{n}} \left\{ y^{\textrm{T}}e : \sum_{i=1}^{n}C(x_{i},e_{i},\lambda)=\sum_{i=1}^{n}C(x_{i},e_{oi},\lambda)\right\} .\label{eq:IDR}
\end{equation}
Problem (\ref{eq:IDR}) describes the assignment of $e$ values to
$y$ values in a sample generated according to a location-scale model,
and it admits $e=e_{o}$ as its only solution. Since $e_{o}$ and
$\lambda$ are unknown, the assignment problem (\ref{eq:IDR}) is
infeasible: we thus introduce the equivalent, feasible formulation
\[
%\textrm{(D)}\quad\max_{e\in\mathbb{R}^{n}}\;\{y^{\textrm{T}}e\mid\begin{cases}
%\sum_{i=1}^{n}x_{i}e_{i} & =0\\
%\frac{1}{2}\sum_{i=1}^{n}x_{i}(e_{i}^{2}-1) & =0
%\end{cases},
\textrm{(D)}\quad\max_{e\in\mathbb{R}^{n}}\left\{ y^{\textrm{T}}e : \sum_{i=1}^{n}x_{i}e_{i} =0, \; \frac{1}{2}\sum_{i=1}^{n}x_{i}(e_{i}^{2}-1) =0\right\},
\]
the dual regression program.

\subsection{Solving the dual program\label{sub:The-Dual-Problem}}

The solution to (D) is easily found from
the Lagrangian 
\[
\mathscr{L}=\sum_{i=1}^{n}y_{i}e_{i}-\lambda_{1}\sum_{i=1}^{n}x_{i}e_{i}-\frac{1}{2}\lambda_{2}\sum_{i=1}^{n}x_{i}(e_{i}^{2}-1).
\]
Differentiating with respect to $e_{i},$ we obtain $n$ first-order
conditions: 
\begin{equation*}
\frac{\partial\mathscr{L}}{\partial e_{i}}=y_{i}-\lambda_{1}^\textrm{T} x_{i}-(\lambda_{2}^\textrm{T} x_{i})e_{i}=0 \quad(i=1,\ldots,n).
\end{equation*}
%Keeping in mind that $x_{i}$ and the Lagrange multipliers $\lambda_{1}$
%and $\lambda_{2}$ are $K$ component vectors, we obtain for each
%$e_{i}$: 
Upon rearranging we obtain the closed-form solution
\begin{equation*}
%e_{i}=\frac{y_{i}-\lambda_{1}^\textrm{T} x_{i}}{\lambda_{2}^\textrm{T} x_{i}} \equiv e(y_{i},x_{i},\lambda) \quad (i=1,\ldots,n),\label{eq:ei}
e_{i}=\frac{y_{i}-\lambda_{1}^\textrm{T} x_{i}}{\lambda_{2}^\textrm{T} x_{i}} \quad (i=1,\ldots,n),
\end{equation*}
which is of the location-scale form $e_{i}=\{y_{i}-\mu(x_{i})\}/\sigma(x_{i})$,
with %the functions 
$\mu(x_{i})$ and $\sigma(x_{i})$ linear
in $x_{i}$. 

Another view is obtained by writing the first-order conditions as
\begin{equation}
y_{i}=\lambda_{1}^\textrm{T} x_{i}+(\lambda_{2}^\textrm{T} x_{i})e_{i} \quad(i=1,\ldots,n),\label{eq:8}
\end{equation}
a linear location-scale representation, with corresponding quantile
regression representation
\begin{equation}
y_{i}=(\lambda_{1}+\lambda_{2}e_{i})^\textrm{T} x_{i}=\{\lambda_{1}+\lambda_{2}F_{n}^{-1}(u_{i})\}^\textrm{T} x_{i}\equiv\beta(u_{i})^\textrm{T} x_{i}\quad(i=1,\ldots,n).\label{eq:9}
\end{equation}
%Thus, the dual regression program (D) 
Program (D) thus provides a complete characterization
of linear representations of the form (\ref{eq:8}) and (\ref{eq:9}),
as they arise from its first-order conditions. Moreover, the parameters
of these representations are the Lagrange multipliers $\lambda_{1}$
and $\lambda_{2}$ of an optimization problem with solution %the $n$-vector
$e=e_{o}$.

The quantile regression representation of the first-order conditions
of (D) sheds additional light on the monotonicity property of dual
regression solutions, when there are no repeated $X$ values. For
$u,u'\in(0,1)$, $u'>u$, the no-crossing property of conditional
quantiles requires 
\[
\beta(u')^\textrm{T} x_{i}-\beta(u)^\textrm{T} x_{i}>0,\quad(i=1,\ldots,n),
\]
which is satisfied if $\lambda_{2}^\textrm{T} x_{i}$ is strictly positive
for each $i$, and coincides with the $n$ second-order conditions
of program (D): 
\[
\frac{\partial^{2}\mathscr{L}}{\partial e_{i}\partial e_{i}}=-\lambda_{2}^\textrm{T} x_{i}<0,\quad(i=1,\ldots,n).
\]
Therefore, an optimal $e$ solution that violates the monotonicity
property is ruled out by the requirement that for an observation with
$X$ value $x_{i}$, the ordering of the $Y$ values $\beta(u')^\textrm{T} x_{i}$
and $\beta(u)^\textrm{T} x_{i}$ must correspond to the ordering of the
$u$ values. Hence the correlation criterion of system (D) suffices
to impose monotonicity, with optimality of a solution then being equivalent
to monotonicity at the $n$ sample points. %Although monotonicity is not imposed, it defines an optimal solution: 
Dual regression is thus able to incorporate this property in the estimation procedure, 
which facilitates extrapolation beyond the empirical support of $X$, and 
yields significant finite-sample improvements in the 
estimation of conditional quantile functions as illustrated by our simulations in $\mathsection$\ref{subsec:simulations}. 
%Moreover, since the crossing point is then located outside the empirical support of $X$, 
%dual regression based conditional quantile functions are also easier to extrapolate.

\subsection{Formal duality\label{sub:Formal-Duality}}

By Lagrangian duality arguments (\citealp{BV:2004}, Chapter 5), the objective function of the dual of problem (D) is 
\[
Q_{n}(\lambda)=\sup_{e\in\mathbb{R}^{n}}y^{\textrm{T}}e-\sum_{i=1}^{n}\left\{ C\left(x_{i},e_{i},\lambda\right)-C\left(x_{i},e_{oi},\lambda\right)\right\} ,
\]
defined for all $\lambda\in\Lambda_{0}$, where $\Lambda_{0}=\Lambda_{1}\times\Lambda_{2}$,
with $\Lambda_{1}=\mathbb{R}^{K}$ and $\Lambda_{2}=\{\lambda_{2}\in\mathbb{R}^{K}:\;\inf_{i\leq n}\lambda_{2}^\textrm{T} x_{i}>0\}$.
Under the conditions of Theorem \ref{Formal duality} below, $Q_{n}(\lambda)$
has a closed-form expression, is strictly convex over $\Lambda_{0}$,
and minimizing $Q_{n}(\lambda)$ over $\Lambda_{0}$ is equivalent
to solving (D). Given a vector $\omega\in\mathbb{R}^{n}$, we let $\textrm{diag}(\omega_{i})$  denote
the $n\times n$ diagonal matrix with diagonal elements  $\omega_{1},\ldots,\omega_{n}$. 

\begin{sloppy}
\begin{condition}The random variable $Y$ is continuously distributed conditional on
$X$, with conditional density $f_{Y\mid X}(y\mid X)$ bounded away from
$0$. \label{Data}\end{condition} 

\begin{condition}%For a specified nonsingular $n\times n$ diagonal
%matrix $\Omega_{n}$ with diagonal elements $\omega_{1},\ldots,\omega_{n}$,
For a specified vector $\omega\in\mathbb{R}^{n}$, the matrix $\textrm{diag}(\omega_{i})$ is nonsingular and 
the matrix $\sum_{i=1}^{n}\omega_{i}^{-1}x_{i}x_{i}^{\textrm{T}}=M_{n}$
is finite, positive definite, and has rank $K$.\label{Design}\end{condition}

\begin{condition}There exists $(\lambda,e_{o})\in\varLambda_{0}\times\mathbb{R}^{n}$
such that $y_{i}=\lambda_{1}^\textrm{T} x_{i}+(\lambda_{2}^\textrm{T} x_{i})e_{oi}$
with $\inf_{i\leq n}\lambda_{2}^\textrm{T} x_{i}\geq\tau$ for some constant
$\tau>0$, and $\sum_{i=1}^{n}x_{i}e_{oi}=0$ and $\sum_{i=1}^{n}x_{i}(e_{oi}^{2}-1)=0$.\label{ExistenceDR}\end{condition}
\par\end{sloppy}

%Given a vector $\omega\in\mathbb{R}^{n}$, we let $\textrm{diag}(\omega_{i})$  denote
%the $n\times n$ diagonal matrix with diagonal elements  $\omega_{1},\ldots,\omega_{n}$. 
Theorem \ref{Formal duality} summarizes our finite-sample
analysis of dual regression. The proofs of all formal results in the
paper are given in the Supplementary Material.
\begin{theorem}
\label{Formal duality}If Conditions \ref{Data}--\ref{ExistenceDR}
hold with %$\Omega_{n}=\textrm{diag}(\lambda_{2}^\textrm{T} x_{i})$, 
$\omega = (\lambda_{2}^\textrm{T} x_{1}, \ldots, \lambda_{2}^\textrm{T} x_{n})$, for
all $\lambda_{2}\in\Lambda_{2}$, then problem (\ref{eq:IDR})
admits the equivalent feasible formulation (D), with solution and
multipliers $e^{*}$ and $\lambda^{*}$, respectively.
Moreover, for program (D) the following holds:% with $e(y_{i},x_{i},\lambda)$ defined in (\ref{eq:ei}):

(i) Primal problem: the dual of (D) is 
\[
%\textrm{(P)}\quad\min_{\lambda\in\Lambda_{0}}\sum_{i=1}^{n}\frac{1}{2}\left\{ \left(\frac{y_{i}-\lambda_{1}^\textrm{T} x_{i}}{\lambda_{2}^\textrm{T} %x_{i}}\right)^{2}+1\right\} \left(\lambda_{2}^\textrm{T} x_{i}\right),
%\textrm{(P)}\quad\min_{\lambda\in\Lambda_{0}}\sum_{i=1}^{n}\frac{1}{2}\left\{ e(y_{i},x_{i},\lambda)^{2}+1\right\} \left(\lambda_{2}^\textrm{T} x_{i}\right),
\textrm{(P)}\quad\min_{\lambda\in\Lambda_{0}}\sum_{i=1}^{n}\frac{1}{2}\left\{\left( \frac{y_{i}-\lambda_{1}^\textrm{T} x_{i}}{\lambda_{2}^\textrm{T} x_{i}}\right)^{2}+1\right\} \left(\lambda_{2}^\textrm{T} x_{i}\right),
\]
the primal dual regression problem, with solution $\lambda_{n}$.

(ii) First-order conditions: program (D) admits the method-of-moments
representation 
\begin{equation}
\sum_{i=1}^{n}x_{i}\left(\frac{y_{i}-\lambda_{1}^\textrm{T} x_{i}}{\lambda_{2}^\textrm{T} x_{i}} \right) = 0,\quad \frac{1}{2}\sum_{i=1}^{n}x_{i}\left\{ \left( \frac{y_{i}-\lambda_{1}^\textrm{T} x_{i}}{\lambda_{2}^\textrm{T} x_{i}} \right)^{2}-1\right\} = 0, \label{eq:DR-MM}
\end{equation}
%\begin{eqnarray}
%\sum_{i=1}^{n}x_{i}e_{i} & = & 0\nonumber \\
%\frac{1}{2}\sum_{i=1}^{n}x_{i}\left(e_{i}^{2}-1\right) & = & 0\label{eq:DR-MM}\\
%e_{i} & = & \frac{y_{i}-\lambda_{1}^\textrm{T} x_{i}}{\lambda_{2}^\textrm{T} x_{i}}\quad(i=1,\ldots,n),\nonumber 
%\end{eqnarray}
the first-order conditions of (P).

(iii) With probability 1: (a) uniqueness: the pair $(\lambda_{n},e^{*})$
is the unique optimal solution to (P) and (D), and $\lambda_{n}=\lambda^{*}$;
(b) strong duality: the value of (D) equals the value of (P).
\end{theorem}

Theorem \ref{Formal duality} establishes formal duality of our initial
assignment problem under first and second moment orthogonality constraints
and the global $M$-estimation problem (P). Convexity of (P) guarantees
that to a unique assignment of $e$ values corresponds a unique linear
representation of the form (\ref{eq:loc-scale mod}). Uniqueness further
implies that if $e_{o}$ satisfies independence, then the orthogonality
conditions in (\ref{eq:DR-MM}) are both necessary and sufficient
for the dual solution $e^{*}$ to satisfy independence. 

The primal problem (P) is a locally heteroscedastic generalization
of a simultaneous location-scale estimator proposed by \citet{Huber:1981}
and further analyzed in \citet{Owen:2001}. The linear heteroscedastic
model of equation (\ref{eq:8}) has been previously encountered in
the quantile regression literature: see \citet{Koenker:Zhao:1994} and \citet{He:1997}.
The former consider the efficient estimation of (\ref{eq:8}) via
$L$-estimation while the latter develops a restricted quantile regression
method that prevents quantile crossing. Compared to these quantile-based
methods, dual regression trades local estimation and the convenient
linear programming formulation of quantile regression for simultaneous
estimation of location and scale parameters.

\subsection{Connection with the dual formulation of quantile regression\label{sub:ConnectionwithQR}}

The dual problem of the linear $0\cdot5$ quantile regression of
$Y$ on $X$ is (cf., Koenker, 2005, p. 87, equation 3.12):
\begin{equation}
\max_{u} \left\{ y^{\textrm{T}}u : \sum_{i=1}^{n}x_{i}\left(u_{i}-\frac{1}{2}\right)=0,\quad u\in[0,1]^{n}\right\} .\label{eq:1}
\end{equation}
The solution to problem (\ref{eq:1}) produces values of $u$ that
are largely 0 and 1, with $K$ sample points being assigned $u$ values
that are neither 0 nor 1. The points that are assigned 1 fall above
the median quantile regression; the points receiving 0's fall below;
and the remaining points fall on the median quantile regression plane.
One direction of extension of (\ref{eq:1}) is to replace the 1/2
with values $\alpha$ that fall between 0 and 1 to obtain the $\alpha$
quantile regression.

Another extension is to augment problem (\ref{eq:1}) by adding $K$
more constraints: 
\begin{equation}
\max_{u}\left\{y^{\textrm{T}}u :
\sum_{i=1}^{n}x_{i}\left(u_{i}-\frac{1}{2}\right) =0,
\;\sum_{i=1}^{n}x_{i}\left(u_{i}^{2}-\frac{1}{3}\right) =0,
\; u\in[0,1]^{n}\right\}.\label{eq:2}
\end{equation}
It is apparent that the solution to (\ref{eq:1}) does not satisfy (\ref{eq:2}):
the variance of $u$ around 0 in the solution to (\ref{eq:1}) is
approximately $1/2$, not $1/3$. To satisfy program (\ref{eq:2}),
the $u$'s have to be moved off $0$ and $1$. Since $x_{i}$ contains
an intercept, the sample moments of $u$ and $u^{2}$ will be $1/2$
and $1/3$; $u$ and $u^{2}$ will be orthogonal to the columns of
the matrix $(x_{1},\ldots,x_{n})^{\textrm{T}}$, relations that are
necessary but not sufficient for uniformity and independence.

Both systems (\ref{eq:1}) and (\ref{eq:2}) demand monotonicity by
maximally correlating $y$ and $u$. %It is worth noting that a
A violation of monotonicity requires there to be two observations that share the
same $X$ values but have different $y$ values, with the lower of
the two $y$ values having the weakly higher value of $u$. However, 
a solution characterized by such a violation could be improved upon
by exchanging the $u$ assignments. Thus violation of monotonicity in program (\ref{eq:1}) 
arises because the set of admissible exchanges in $u$ assignments is overly restricted: 
(\ref{eq:1}) is dual to a linear program
well-known to have solutions at which $K$ observations are interpolated
when $K$ parameters are being estimated, i.e., the hyperplanes obtained
by regression quantiles must interpolate $K$ observations. 

By reformulating program (\ref{eq:2}) into a constrained optimization
problem over $\mathbb{R}^{n}$, program (D) further expands the set
of admissible exchanges in $u$ assignments, since $u$ is restricted
to $[0,1]^{n}$. Doing this, the problem corresponding to (\ref{eq:2})
becomes the dual regression program (D), where $e$ can take on any
real value. It is then natural to take $u_{i}^{*}=F_{n}(e_{i}^{*}),$
the empirical cumulative distribution function of the dual regression
solution $e^{*}$, thereby imposing uniformity to high precision even
at small $n$.

\section{Generalization}\label{sec:Generalization}

\subsection{Infeasible generalized dual regression}

The dual regression characterization of location-scale conditional
distribution functions via the monotonicity element, the objective,
and the independence element, the constraints, can be exploited to
characterize more flexible representations. Similarly to the approach
introduced in $\mathsection$\ref{sec:Basics}, we first analyze the infeasible
assignment problem for a general representation of the stochastic
structure of $Y$ conditional on $X$:
\begin{equation}
Y=H(X,\varepsilon)\equiv H_{X}(\varepsilon),\quad\varepsilon \mid X\sim F,\label{eq:GDRrep}
\end{equation}
where $F$ is a specified cumulative distribution function with support
the real line, and for each value $x$ of $X$, the derivative $H_{x}'(\varepsilon)$
of $H_{x}(\varepsilon)$ is strictly positive. Representation (\ref{eq:GDRrep})
always exists with $H_{x}$ defined as the composition of the conditional
quantile function of $Y$ given $X=x$ and the distribution function
$F$.

To each monotone function $H_{x}$ also corresponds a convex function
$\widetilde{H}_{x}$ defined as 
\[
\widetilde{H}_{x}(e)\equiv\int_{0}^{e}H_{x}(s)ds \quad(e\in\mathbb{R}).
\]
The monotonicity of $H_{x}(\varepsilon)$ guarantees the convexity
of $\widetilde{H}_{x}(\varepsilon)$. The slope of this function gives
the value of $Y$ corresponding to a value $e$ of $\varepsilon$
at $X=x$. Thus $F_{Y\mid X}(Y\mid X)$ corresponds to a collection of
convex functions, with one element of this collection for each value
of $X,$ together with a single random variable whose distribution
is common to all the convex functions.

Equipped with $\widetilde{H}_{X}$, suppose we are tasked with assigning
a value $e_{i}$ to each of the $n$ realizations $\{(y_{i},x_{i})\}_{i=1}^{n}$.
Then, for $S_{n}=\sum_{i=1}^{n}\widetilde{H}_{x_{i}}(\varepsilon_{i})$,
solving the infeasible problem 
\[
\textrm{(IGD)} \quad \max_{e\in\mathbb{R}^{n}} \left\{ y^{\textrm{T}}e : \sum_{i=1}^{n}\widetilde{H}_{x_{i}}(e_{i})=S_{n}\right\} ,
\]
generates the correct $y-e$ assignment: writing the Lagrangian
\[
\mathscr{L}=y^{\textrm{T}}e-\Lambda\left\{ \sum_{i=1}^{n}\widetilde{H}_{x_{i}}(e_{i})-S_{n}\right\} ,
\]
the $n$ associated first-order conditions are 
\begin{equation}
\frac{\partial\mathscr{L}}{\partial e_{i}}=y_{i}-\Lambda H_{x_{i}}(e_{i})=0 \quad(i=1,\ldots,n),\label{eq:FOCs17n}
\end{equation}
\begin{sloppy}and convexity of $\widetilde{H}_{x_{i}}$ then guarantees
that (\ref{eq:FOCs17n}) is uniquely satisfied by $(\Lambda,e)=(1,\varepsilon_{o})$,
with $\varepsilon_{o}=(\varepsilon_{1},\ldots,\varepsilon_{n})^{\textrm{T}}$.
This demonstrates that maximizing $y^{\textrm{T}}e$ generally suffices
to match $e$'s to $y$'s, regardless of the form of $H_{X}$
in (\ref{eq:GDRrep}).

\begin{theorem}
Suppose that (\ref{eq:GDRrep}) holds with $H_{x_{i}}:\mathbb{R}\rightarrow\mathbb{R}$
a continuously differentiable function that satisfies $\inf_{e\in\mathbb{R}}H'_{x_{i}}(e)\geq\tau$
for each $x_{i}$ and some constant $\tau>0$. Then the infeasible
generalized dual regression problem (IGD) with $S_{n}=\sum_{i=1}^{n}\widetilde{H}_{x_{i}}(\varepsilon_{i})$
generates the correct $y-e$ assignment, i.e., the pair $(\Lambda,e)=(1,\varepsilon_{o})$
uniquely solves first-order conditions (\ref{eq:FOCs17n}).\label{thm:IGD} 
\end{theorem}

Theorem \ref{thm:IGD} shows that  problem (IGD) fully
characterizes the $y-e$ assignment problem: given $\widetilde{H}_{x_{i}}$
and $S_{n}$, solving (IGD) assigns a value $e_{i}$ to
each sample point $(y_{i},x_{i})$, and this value is the corresponding
value $F_{Y\mid X}(y_{i}\mid x_{i})$ up to a specified transformation
$F$. If $F$ is specified to be a known distribution, the $n$ values
$F_{Y\mid X}(y_{i}\mid x_{i})$ are then also known. If $F$ is specified
to be an unknown distribution, as in our application below, the empirical
distribution of $\varepsilon_{o}$ then provides an asymptotically
valid estimator for $F$. Knowledge of $\widetilde{H}_{x_{i}}$ and
$S_{n}$ can thus be incorporated into a mathematical programming
problem which delivers the values of $F_{Y\mid X}$ at the $n$ sample
points.\end{sloppy}

\subsection{Generalized dual regression representations: definition and characterization}

Problem (IGD) is infeasible because neither $\widetilde{H}_{x_{i}}$
nor $S_{n}$ is known. However, Theorem \ref{thm:IGD} motivates
a feasible approach once $H_{X}$ and $F$ are specified. Denote the
components of $X$ without the intercept by $\widetilde{X}$, so that
$X=(1,\widetilde{X})^{\textrm{T}}$. Without loss of generality, let $\widetilde{X}$
be centered, denoted $\widetilde{X}^{c}$, and let $X^{c}=(1,\widetilde{X}^{c})^{\textrm{T}}$.
With $h_{1}(\varepsilon)=1$ and $h_{2}(\varepsilon)=\varepsilon$,
we specify %each of the monotone functions $H_{x}$ 
$H_{X}$ by a linear combination
of $J$ basis functions $h(\varepsilon)=\{h_{1}(\varepsilon),\ldots,h_{J}(\varepsilon)\}^{\textrm{T}}$,
the coefficients of which depend on $X$: 
\begin{equation}
H_{X}(\varepsilon)=\sum_{j=1}^{J}\beta_{j}(X)h_{j}(\varepsilon),\label{eq:Hfunc}
\end{equation}
and we assume that $H_{X}$ is linear in $X$ and set: 
\begin{equation}
\beta_{j}(X)=\alpha_{j}+\beta_{j}^\textrm{T}\widetilde{X}^{c} \quad(j=1,\ldots,J).\label{eq:22}
\end{equation}
Finally, we specify a zero mean and unit variance distribution for
$\varepsilon$ by imposing $E(\varepsilon)=0$ and $E(\varepsilon^{2}-1)/2=0$,
and setting $\alpha_{j}=0$ for $j=3,\ldots,J$, in (\ref{eq:22}).

With $\alpha_{2}+\beta_{2}^\textrm{T}\widetilde{X}^{c}>0$, our normalization
and (\ref{eq:Hfunc})--(\ref{eq:22}) together yield the augmented,
generalized dual regression model
\begin{equation}
Y=\alpha_{1}+\alpha_{2}\varepsilon+\beta_{1}^\textrm{T}\widetilde{X}^{c}+(\beta_{2}^\textrm{T}\widetilde{X}^{c})\varepsilon+\sum_{j=3}^{J}(\beta_{j}^\textrm{T}\widetilde{X}^{c})h_{j}(\varepsilon)\equiv H_{X}(\varepsilon;\alpha,\beta),\quad\varepsilon\mid X\sim F.\label{eq:DGP aug loc-scale}
\end{equation}
Equation (\ref{eq:DGP aug loc-scale}) admits of the following interpretation.
When $\widetilde{X}^{c}=0,$ $Y=\alpha_{1}+\alpha_{2}\varepsilon$ and
$\varepsilon=(Y-\alpha_{1})/\alpha_{2}$, so that $\varepsilon$ is
just a re-scaled version of the distribution of $Y$ at $\widetilde{X}^{c}=0$.
Since $\varepsilon$ is independent of $X$, transformations of this
shape of $\varepsilon$ must suffice to produce $Y$ at other values
of $X$. The first two transformations, $\beta_{1}^\textrm{T}\widetilde{X}^{c}$
and $(\beta_{2}^\textrm{T}\widetilde{X}^{c})\varepsilon$, are translations
of location and scale which do not essentially affect the shape
of $Y$'s response to changes in $\varepsilon$ at all. The additional
terms $(\beta_{j}^\textrm{T}\widetilde{X}^{c})h_{j}(\varepsilon)$ achieve that
end.

Suppose that we observe a sample of $n$ identically and independently
distributed realizations $\{(y_{i},x_{i})\}_{i=1}^{n}$ generated
according to model (\ref{eq:DGP aug loc-scale}). Define $x_{ij}^{c}=x_{i}^{c}$ for $j=1,2$, and
$x_{ij}^{c}=\widetilde{x}_{i}^{c}$ for $j=3,\ldots,J$, and let $(\gamma,\lambda)\in\mathbb{R}^{2+J(K-1)}$
and $e_{o}\in\mathbb{R}^{n}$ satisfy the system of $n$ equations
and $2n$ inequality constraints
%\begin{equation}
%y_{i}=\gamma_{1}+\gamma_{2}e_{oi}+(\lambda_{1}^\textrm{T}\widetilde{x}_{i}^{c})+(\lambda_{2}^\textrm{T}\widetilde{x}_{i}^{c})e_{oi}+\sum_{j=3}^{J}
%(\lambda_{j}^\textrm{T}\widetilde{x}_{i}^{c})h_{j}(e_{oi}) \quad(i=1,\ldots,n),\label{eq:aug loc-scale mod}
%\end{equation}
%with $\gamma_{2}+(\lambda_{2}^\textrm{T}\widetilde{x}_{i}^{c})>0$ and $H'_{x_{i}}(e_{oi};\gamma,\lambda)>0$,
\begin{equation}
y_{i}=H_{x_{i}}(e_{oi};\gamma,\lambda), \quad \gamma_{2}+\lambda_{2}^\textrm{T}\widetilde{x}_{i}^{c}>0, \quad H'_{x_{i}}(e_{oi};\gamma,\lambda)>0 \quad (i=1,\ldots,n),\label{eq:aug loc-scale mod}
\end{equation}
where $e_{o}$ further satisfies $\sum_{i=1}^{n}x_{ij}^{c}\widetilde{h}_{j}(e_{oi})=0$ $(j=1,\ldots,J)$, with 
$\widetilde{h}_{1}(e_{oi})=e_{oi}$, $\widetilde{h}_{2}(e_{oi})=(e_{oi}^{2}-1)/2$,
and $\widetilde{h}_{j}(e_{oi})=\int_{0}^{e_{oi}}h_{j}(s)ds$
$(j=3,\ldots,J)$. These relations reduce to the linear heteroscedastic
representation of $\mathsection$\ref{sec:Basics} for $J=2$, and impose that $e_0$ be a zero mean and unit variance vector satisfying the augmented set of orthogonality
conditions $\sum_{i=1}^{n}\widetilde{x}_{i}^{c}\widetilde{h}_{j}(e_{oi})=0$ $(j=1,\ldots,J)$. The sequence of
vectors $e_{o}$ that satisfies the generalized dual regression
representation (\ref{eq:aug loc-scale mod}) as well as the associated
orthogonality constraints for each $n$ then provides an asymptotically
valid characterization of the data-generating process (\ref{eq:DGP aug loc-scale}).

Each element of this sequence is characterized by the assignment problem
\begin{equation}
\max_{e\in\mathbb{R}^{n}} \left\{ y^{\textrm{T}}e : 
\sum_{i=1}^{n}\widetilde{H}_{x_{i}}(e_{i};\theta)=\sum_{i=1}^{n}\widetilde{H}_{x_{i}}(e_{oi};\theta)\right\},\label{eq:IGD-1}
\end{equation}
where $\widetilde{H}_{x_{i}}(e_{i};\theta)=\int_{0}^{e_{i}}H_{x_{i}}(s;\theta)ds$,
and $\theta=(\theta_{1},\ldots,\theta_{J})^{\textrm{T}}$, with $\theta_{j}=(\gamma_{j},\lambda_{j})^{\textrm{T}}\in\mathbb{R}^{K}$
for $j=1,2$, and $\theta_{j}=\lambda_{j}\in\mathbb{R}^{K-1}$ for
$j=3,\ldots,J$. 
%Letting $x_{ij}^{c}=x_{i}^{c}$ for $j=1,2$ and
%$x_{ij}^{c}=\tilde{x}_{i}^{c}$ for $j=3,\ldots,J$, 
%the orthogonality
%conditions satisfied by $e_{o}$ can be written in the compact form
%$\sum_{i=1}^{n}x_{ij}^{c}\tilde{h}_{j}(e_{oi})=0$, $j=1,\ldots,J$.
%With this notation in hand, 
Since $e_{0}$ and $\theta$ are unknown, problem (\ref{eq:IGD-1}) is infeasible; we thus formulate an equivalent, feasible
implementation of problem (IGD):
\[
\textrm{(GD)}\quad\max_{e\in\mathbb{R}^{n}} \left\{ y^{\textrm{T}}e : \sum_{i=1}^{n}x_{ij}^{c}\widetilde{h}_{j}(e_{i})=0\quad(j=1,\ldots,J)\right\} ,
\]
the generalized dual regression program. (GD) then uniquely characterizes
representation (\ref{eq:aug loc-scale mod}).

In order to state the properties of (GD) formally, we define the
parameter space $\Theta_{n}$, which specifies parameter values compatible
with monotone representations: 
\[
\Theta_{n}=\left\{ \theta\in\Theta_{0,n}:\textrm{there exists }e\in\mathbb{R}^{n}:y_{i}=H_{x_{i}}(e_{i};\theta)\textrm{ and }\inf_{e\in\mathbb{R}}H'_{x_{i}}(e;\theta)>0\quad(i=1,\ldots,n)\right\} ,
\]
with $\Theta_{0,n}=\{\theta\in\mathbb{R}^{2+J(K-1)}:\;\inf_{i\leq n}\theta_{2}^\textrm{T} x_{i}^{c}>0\}$.
For $\theta\in\Theta_{n}$, let $e(y_{i},x_{i},\theta)$ denote the
inverse function of $H_{x_{i}}(e_{i};\theta)$, which is well-defined
for each $x_{i}$. %, and let $\phi(\theta)=[H_{x_{1}}'\{e(y_{1},x_{1},\theta);\theta\},\dots,H_{x_{n}}'\{e(y_{n},x_{n},\theta);\theta\}]^{\textrm{T}}$. 
We assume that the basis functions $h$ and the pair $(\theta,e_{0})$ satisfy
the following conditions.

\begin{sloppy}
\begin{condition}There exists a finite constant $C_{h}$ such that
$\max_{j=3,\ldots,J}\sup_{e\in\mathbb{R}}\{|h_{j}(e)|+|\widetilde{h}_{j}(e)|\}\leq C_{h}$,
and the matrix $E[h\{e(Y,X,\theta)\}h\{e(Y,X,\theta)\}^{\textrm{T}} \mid X=x_{i}]$
is finite and nonsingular for each $x_{i}$ and all $\theta\in\Theta_{n}$.\label{FR}\end{condition}

\begin{condition}There exists $(\theta,e_{o})\in\Theta_{n}\times\mathbb{R}^{n}$
such that $y_{i}=H_{x_{i}}(e_{oi};\theta)$ and $\inf_{e\in\mathbb{R}}H'_{x_{i}}(e;\theta)\geq\tau$,
for $i=1,\ldots,n$ and some constant $\tau>0$, and $e_{o}$ satisfies
$\sum_{i=1}^{n}x_{ij}^{c}\widetilde{h}_{j}(e_{oi})=0$, for $j=1,\ldots,J$.\label{ExistenceGDR}\end{condition}
\par\end{sloppy}

Let $\phi(\theta)=[H_{x_{1}}'\{e(y_{1},x_{1},\theta);\theta\},\dots,H_{x_{n}}'\{e(y_{n},x_{n},\theta);\theta\}]^{\textrm{T}}$. Theorem \ref{thm:Formal duality-GDR} summarizes our finite-sample
analysis of generalized dual regression.

\begin{theorem}\label{thm:Formal duality-GDR}
If Conditions \ref{Data}, \ref{Design}, \ref{FR} and \ref{ExistenceGDR} hold with 
%$\Omega_{n}=\textrm{diag}[H_{x_{i}}'\{e(y_{i},x_{i},\theta);\theta\}]$,
%$\omega=[H_{x_{1}}'\{e(y_{1},x_{1},\theta);\theta\},\dots,H_{x_{n}}'\{e(y_{n},x_{n},\theta);\theta\}]$,
$\omega=\phi(\theta)$,
for all $\theta\in\Theta_{n}$, then problem (IGD)
admits the equivalent feasible formulation (GD), with solution and
multipliers $e^{*}$ and $\theta^{*}$, respectively. Moreover,
for program (GD) the following holds:

(i) Primal problem: the dual of (GD) is
%\[
%\textrm{(GP)}\quad\min_{\theta\in\Theta_{n}}\left\{ \sum_{i=1}^{n}\sum_{j=2}^{J}(\theta_{j}^\textrm{T} x_{ij}^{c})\{h_{j}(e_{i})e_{i}-\widetilde{h}_{j}(e_{i})\} : y_{i}=\sum_{j=1}^{J}(\theta_{j}^\textrm{T} x_{ij}^{c})h_{j}(e_{i}) \quad (i=1,\ldots,n)\right\} ,
%\]
\begin{equation*}
\textrm{(GP)}\quad\min_{\theta\in\Theta_{n}}  \sum_{i=1}^{n}\sum_{j=2}^{J}(\theta_{j}^\textrm{T} x_{ij}^{c})\left[h_{j}\left\{e(y_{i},x_{i},\theta)\right\}e(y_{i},x_{i},\theta)-\widetilde{h}_{j}\left\{e(y_{i},x_{i},\theta)\right\} \right],
\end{equation*}
the primal generalized dual regression problem, with solution $\theta_{n}$.

(ii) First-order conditions: program (GD) admits the method-of-moments
representation 
%\begin{eqnarray}
%\sum_{i=1}^{n}x_{ij}^{c}\widetilde{h}_{j}(e_{i})=0\quad(j=1,\ldots,J) & \textrm{and} & y_{i}=\sum_{j=1}^{J}(\theta_{j}^\textrm{T} x_{ij}^{c})h_{j}(e_{i})\quad(i=1,\ldots,n),%\end{eqnarray}
\begin{equation}
\sum_{i=1}^{n}x_{ij}^{c}\widetilde{h}_{j}\left\{e(y_{i},x_{i},\theta)\right\}=0 \quad(j=1,\ldots,J),\label{eq:GDR_MM}
\end{equation}
the first-order conditions of (GP).

(iii) With probability 1: (a) uniqueness: the pair $(\theta_{n},e^{*})$
is the unique optimal solution to (GP) and (GD), and $\theta_{n}=\theta^{*}$;
(b) strong duality: the value of (GD) equals the value of (GP).
\end{theorem}

Problem (GD) augments the set of orthogonality constraints in (D)
and generates increasingly flexible representations of the form (\ref{eq:aug loc-scale mod}).
For each element of this sequence, (GD) then provides a feasible formulation
of the generalized $y-e$ assignment problem (IGD)
with optimality condition $-H'_{x_{i}}(e_{i}^{*};\theta^{*})<0$ equivalent
to monotonicity. Theorem \ref{thm:Formal duality-GDR} also states
the form of the corresponding primal problem, whose convexity guarantees
that (GD) and (GP) uniquely and equivalently characterize representation
(\ref{eq:DGP aug loc-scale}). Uniqueness further implies that if
$e_{o}$ satisfies independence, then the orthogonality conditions
in (\ref{eq:GDR_MM}) are both necessary and sufficient for the dual
solution $e^{*}$ to satisfy independence as well. Theorem \ref{thm:Formal duality-GDR} 
thus characterizes and establishes the duality between specification of
orthogonality constraints on $e$ and specification of a globally
monotone representation for $Y$ conditional on $X$.

Formally, (GP) is the restriction of the dual of (GD) to $\Theta_{n}$.
The existence Condition \ref{ExistenceGDR} and the form of (GD) optimality
conditions together ensure that (GD) does not admit a global maximum
with associated multipliers outside $\Theta_{n}$. Implementing (GP)
thus requires the imposition of inequality constraints with $e_{i}$
only implicitly defined in the specification of (GP) for $J>2$, and
problem (GD)  therefore provides a greatly simplified dual implementation.
The special case of dual regression corresponds to $J=2$, and imposing
$\sum_{i=1}^{n}\widetilde{h}_{j}(e_{i})=0$, for $j=1,2$, is a normalization.
The simple basis $\{e_{i},(e_{i}^{2}-1)/2\}$ is obviously impoverished
for the space of all convex functions, although quite practical for
many applications once the flexibility in the distribution of $e$
is taken into account.

\subsection{Connection with optimal transport formulation of quantile regression}

An alternative approach is to specify $F$ to a known distribution,
and alter representation (\ref{eq:DGP aug loc-scale}) and the corresponding
problem accordingly. If $F$ is specified to be the standard uniform
distribution, then  (\ref{eq:2}) in $\mathsection$\ref{sub:ConnectionwithQR}
can be generalized as 
\begin{equation}
\max_{u\in[0,1]^{n}} \left\{ y^{\textrm{T}}u : \frac{1}{j}\sum_{i=1}^{n}x_{i}\left(u_{i}^{j}-\frac{1}{j+1}\right)=0 \quad (j=1,\ldots,J)\right\}.\label{eq:UniformDR}
\end{equation}
For $u_{i}\in[0,1]$, let $m^{J}(u_{i})=\{m_{J1}(u_{i}),\ldots,m_{JJ}(u_{i})\}^{\textrm{T}}$,
with $m_{Jj}(u_{i})=j^{-1}\{u_{i}^{j}-(j+1)^{-1}\}$. With $\otimes$
denoting the Kronecker product, the large-sample form of program (\ref{eq:UniformDR})
is
\begin{equation}
\max_{U_{J}\in(0,1)} \left\{ E(YU_{J}) : E\{X\otimes m^{J}(U_{J})\}=0\right\}.\label{eq:Population UDR}
\end{equation}
Letting $J$ increase, both the distributional and the orthogonality
constraints get strengthened. Because $X$ includes an intercept,
the distribution of $U_{J}$ approaches uniformity, while simultaneously
satisfying an increasing sequence of orthogonality constraints. In
the limit, a uniformly distributed random variable $U$ satisfying
the full set of orthogonality constraints is thus specified. Since
$E\{X\otimes m^{J}(U)\}=0$ for all $J$ is equivalent to the mean-independence
property $E(X\mid U)=E(X)$ and the uniformity constraint $U\sim U(0,1)$,
in the large $J$ limit program (\ref{eq:Population UDR}) coincides
with the scalar quantile regression problem proposed in independent
work by \citet{CCG:2016} (cf., equation 19, p. 1180) 
\begin{equation}
\max \left\{ E(YU) : U\sim U(0,1),\;E(X\mid U)=E(X)\right\},\label{eq:ScalarQR}
\end{equation}
which provides an optimal transport formulation of quantile regression
(we are grateful to an anonymous referee for highlighting this connection).
Program (\ref{eq:ScalarQR}) is directly amenable to a linear programming
implementation which maintains and exploits the full specification
of the marginal distribution of $U$ to a known distribution, whereas
(\ref{eq:Population UDR}) provides a sequential nonlinear programming
characterization of $U$ which relaxes uniformity for finite $n$
and $J$.

For $e_{i}\in\mathbb{R}$, let $\widetilde{h}^{J}(e_{i})=\{\widetilde{h}_{1}(e_{i}),\ldots,\widetilde{h}_{J}(e_{i})\}^{\textrm{T}}$.
The large-sample form of program (GD) is
\begin{equation}
\max_{e_{J}\in\mathbb{R}}\left\{ E(Ye_{J}) :
E(e_{J})=0, \; E(e_{J}^{2}-1) =0, \;
E\{\widetilde{X}^{c}\otimes\widetilde{h}^{J}(e_{J})\} =0\right\}.\label{eq:PopGDR}
\end{equation}
Program (\ref{eq:PopGDR}) relaxes the support constraint in (\ref{eq:UniformDR})
and only specifies first and second moments of $e_{J}$, while the
centering of $X$ ensures that this is sufficient for $e_{J}$ to
be uniquely determined. The empirical distribution of solutions of
the finite-sample analog (GD) of (\ref{eq:PopGDR}) then provides
an asymptotically valid characterization of the distribution of $e_{J}$.

Letting $J$ increase, orthogonality constraints in (\ref{eq:PopGDR})
are strengthened, and $e_{J}$ gets closer and closer to satisfying
the mean-independence property $E(\widetilde{X}^{c}\mid e_{J})=0$. It follows
that for $J$ large enough, (\ref{eq:PopGDR}) is equivalent to
\begin{equation}
\max_{e\in\mathbb{R}}\left\{E(Ye) :
E(e)=0, \; E(e^{2}-1) =0, \; E(\widetilde{X}^{c}\mid e)  = 0\right\},
\label{eq:limitingGDR}
\end{equation}
the limiting generalized dual regression problem. Theorem \ref{thm:limGDR}
summarizes this discussion.

\begin{sloppy}
\begin{theorem}
Assume that $E(||X||^{2})<\infty$. (i) Suppose that for any $a(U)$
with $E\{a(U)^{2}\}<\infty$ there are $J\times1$ vectors $\psi_{J}$
such that as $J\rightarrow\infty$, $E[\{a(U)-m^{J}(U)^{T}\psi_{J}\}^{2}]\rightarrow0$.
Then programs (\ref{eq:Population UDR}) and (\ref{eq:ScalarQR})
are equivalent in the large $J$ limit. (ii) Suppose that for any
$b(e)$ with $E\{b(e)^{2}\}<\infty$ there are $J\times1$ vectors
$\psi_{J}$ such that as $J\rightarrow\infty$, $E[\{b(e)-\widetilde{h}^{J}(e)^{T}\psi_{J}\}^{2}]\rightarrow0$.
Then programs (\ref{eq:PopGDR}) and (\ref{eq:limitingGDR}) are
equivalent for $J$ large enough.\label{thm:limGDR} 
\end{theorem}
\par\end{sloppy}

\section{Asymptotic Properties}\label{sec:Asymptotics}

We apply our framework to the estimation of a $J$--term 
generalized dual regression model of the form (\ref{eq:DGP aug loc-scale}).
Denote the support of $X$ by $\mathcal{X}$, and, for some finite
constant $C_{\theta}$, define $\Theta_{0}=\{\theta\in\mathbb{R}^{2+J(K-1)}:||\theta||\leq C_{\theta}\,\textrm{and}\:\inf_{x\in\mathcal{X}}\theta_{2}^\textrm{T} x^{c}>0\}$.
Letting $\mathcal{C}^{1}(\mathbb{R})$ denote the space of continuously
differentiable functions on $\mathbb{R}$, define the space of strictly
increasing functions indexed by $X$ values, $\mathcal{M}(X)=\{e_{X}\in\mathcal{C}^{1}(\mathbb{R}):\inf_{y\in\mathbb{R}}e'_{x}(y)>0\;\textrm{for all}\,x\in\mathcal{X}\}$.
The large-sample analog of $\Theta_{n}$ is then the space of vectors
in $\Theta_{0}$ such that there exists a corresponding optimal generalized
dual regression representation: 
\[
\Theta=\left\{ \theta\in\Theta_{0}:\textrm{there exists }e_{X}\in\mathcal{M}(X)\,\textrm{with}\,\Pr[Y=H_{X}\{e_{X}(Y);\theta\}]=1\right\} .
\]
For any $\theta\in\Theta$, denote $e_{X}$ in $\mathcal{M}(X)$ such
that $\Pr[Y=H_{X}\{e_{X}(Y);\theta\}]=1$ by $e(Y,X,\theta)$. 

\begin{condition}For some $\theta_{0}\in\Theta$ and some mean zero
and unit variance cumulative distribution function $F$, the representation
$Y=H_{X}(\varepsilon;\theta_{0})$ holds with probability one, with
$\varepsilon\sim F$ and $E\{\widetilde{X}h_{j}(\varepsilon)\}=0$, for $j=1,\ldots,J$,
and $\inf_{e\in\mathbb{R}}H'_{X}(e;\theta_{0})\geq\tau$ for some
constant $\tau>0$.\label{DGP}\end{condition}

\begin{condition}The matrix $M_{n}$ defined in
Condition \ref{Design} satisfies $\textrm{lim }n^{-1}M_{n}=M$, 
a positive definite matrix of rank $K$, 
and for all $\theta\in\Theta$
the matrix $E[h\{e(Y,X,\theta)\}h\{e(Y,X,\theta)\}^{\textrm{T}}\mid X]$
is nonsingular.\label{LimFR}\end{condition}

%\begin{condition}(i) $E(Y^{2})$, $E(\left\Vert X\right\Vert ^{4})$
%and $E(Y^{2}\left\Vert X\right\Vert ^{2})$ are finite; (ii) $E(Y^{4})$,
%$E(\left\Vert X\right\Vert ^{6})$ and $E(Y^{4}\left\Vert X\right\Vert ^{2})$
%are finite.\label{Moments}\end{condition}
\begin{condition}(i) Let $E(Y^{2})<\infty$, $E(\left\Vert X\right\Vert ^{4})<\infty$
and $E(Y^{2}\left\Vert X\right\Vert ^{2})<\infty$; (ii) let $E(Y^{4})<\infty$,
$E(\left\Vert X\right\Vert ^{6})<\infty$ and $E(Y^{4}\left\Vert X\right\Vert ^{2})<\infty$.\label{Moments}\end{condition}

These conditions are %sufficient for
used to establish existence and consistency of dual regression solutions,
and Condition \ref{Moments}(ii) is needed for asymptotic
normality of estimates of $\theta_{0}$. In view of uniqueness stated
in part (iii) of Theorem \ref{thm:Formal duality-GDR}, these properties are
shared by $\theta_{n}$ and $\theta^{*}$, which we denote by $\widehat{\theta}$
for notational simplicity. We also denote both $e_{i}^{*}$ and indirect
estimates $e(y_{i},x_{i},\theta_{n})$, constructed after solving
(GP), by $\hat{e}_{i}$, with empirical distribution function $F_{n}(e)=n^{-1}\sum_{i=1}^{n}1(\hat{e}_{i}\leq e)$,
$e\in\mathbb{R}$. Furthermore, part (ii) of Theorem \ref{thm:Formal duality-GDR}
shows that while the solution $e^{*}$ is obtained directly by solving
the mathematical program (GD), knowledge that the solution obeys %system
representation (\ref{eq:aug loc-scale mod}) can be exploited to write estimating equations
for $\hat{\theta}$ in the form of system (\ref{eq:GDR_MM}). The computation
of the asymptotic distribution of $\hat{\theta}$ follows from this
characterization.

\begin{theorem}
If $\{(y_{i},x_{i})\}_{i=1}^{n}$ are identically and independently
distributed, and Conditions \ref{Data}, \ref{Design}, \ref{FR},
and \ref{DGP}--\ref{Moments} hold with 
%$\Omega_{n}=\textrm{diag}[H_{x_{i}}'\{e(y_{i},x_{i},\theta);\theta\}]$,
%$\omega=[H_{x_{1}}'\{e(y_{1},x_{1},\theta);\theta\},\dots,H_{x_{n}}'\{e(y_{n},x_{n},\theta);\theta\}]$, 
$\omega=\phi(\theta)$,
for all $\theta\in\Theta_{n}$, then (i) there exists $\hat{\theta}$
in $\Theta$ with probability approaching one, (ii) $\hat{\theta}$
converges in probability to $\theta_{0}$, and (iii) $n^{1/2}(\hat{\theta}-\theta_{0})$
converges in distribution to $N(0,\Sigma)$, with $\Sigma$ defined in (5.6) in the Supplementary Material.\label{thm:AsyThy_Lambda}
\end{theorem}

Knowledge of the statistical properties of $\hat{\theta}$ can be
used to establish the limiting behaviour of the empirical distribution
of $\hat{e}$. Define the empirical dual regression process
\[
\mathbb{U}_{n}(e)=n^{1/2}\{F_{n}(e)-F(e)\}\quad(e\in\mathbb{R}).
\]
Theorem \ref{thm:Weakcv} establishes weak convergence of the empirical
distribution of $\hat{e}$ and the limiting behaviour of $\mathbb{U}_{n}$,
accounting for its dependence on the distribution of $n^{1/2}(\hat{\theta}-\theta_{0})$.

\begin{theorem}
\label{thm:Weakcv}%Suppose that 
If the conditions of Theorem \ref{thm:AsyThy_Lambda}
hold, and, %further assume that 
uniformly in $x$ over $\mathcal{X}$,
$f_{Y\mid X}(y\mid x)$ is uniformly continuous in $y$, bounded and, for
some finite constant $C_{f}$ and all $\theta\in\Theta_{n}$, satisfies
$\sup_{e\in\mathbb{R}}e^{2}f_{Y \mid X}\{H_{x}(e;\theta) \mid x\}\leq C_{f}$, %.
then (i) $\sup_{e\in\mathbb{R}}|F_{n}(e)-F(e)|$ converges in probability
to zero, and (ii) $\mathbb{U}_{n}$ converges weakly to a zero-mean
Gaussian process $\mathbb{U}$ with covariance function defined in
(5.11) in the Supplementary Material.
\end{theorem}

Theorems \ref{thm:AsyThy_Lambda} and \ref{thm:Weakcv} together establish
that the pair $(\hat{\theta},\hat{e})$ provides an asymptotically
valid characterization of the generalized dual regression representation
specified in Condition \ref{DGP}. When $\varepsilon$ is independent
of $X$, Theorem \ref{thm:Weakcv} further implies that the empirical
distribution of $\hat{e}$ provides an asymptotically valid estimator
of  the conditional distribution of $Y$ given $X$. For $u\in(0,1)$,
estimates of the $X$ coefficients in quantile regression form can
then be constructed as $\sum_{j=1}^{J}\hat{\lambda}_{j}h_{j}\{F_{n}^{-1}(u)\}$,
exploiting the structure of the conditional quantile function of $Y$
given $X$ implied by representation (\ref{eq:DGP aug loc-scale}).
Theorem \ref{thm:Weakcv} also establishes asymptotic normality of
the empirical dual regression process. The form of the covariance
function of $\mathbb{U}$ reflects the influence of imposing sample
orthogonality constraints in (GD) on the empirical distribution of
$e^{*}$, or equivalently, of sample variability of parameter estimates
$\theta_{n}$ on the empirical distribution of $e(y_{i},x_{i},\theta_{n})$,
as expected from the classical result of \citet{Durbin:1973}. %Methods
%for inference on parametric empirical processes (see e.g., \citealp{Koenker:Xiao:2002} and 
%\citealp{Parker:2013}) provide a natural direction for future study
%of inference on the empirical dual regression process.

Theorem \ref{thm:Weakcv} can be applied to perform 
pointwise inference on the conditional distribution function of $Y$ conditional on $X$. However, 
simultaneous inference over regions of the joint support of $Y$ and $X$ %outcome and covariate spaces 
is typically of interest in practice. Several approaches 
for uniform inference in the presence of non-pivotal limit processes have been considered in the literature 
(e.g., \citealp{Koenker:Xiao:2002}, and \citealp{Parker:2013}), 
including %resampling and 
simulation methods \citep{CFG:2013}. 
Extension of existing results to dual regression is beyond the scope of this paper but 
they provide a natural direction for future study of uniform inference on the empirical dual regression process.

\section{Engel's Data Revisited}\label{sec:Engel's-Data-Revisited}

\subsection{Empirical analysis}

The classical dataset collected by Engel consists of food expenditure
and income measurements for 235 households, and has been studied by means of quantile regression methods \citep{Koenker:2005}.
We illustrate dual regression methods by estimating the statistical
relationship between food expenditure and income, with household income
as a single regressor and food expenditure as the outcome of interest.

We specify the vector of basis functions by means of trigonometric
series. Alternative choices such as splines and shape-preserving wavelets
%(e.g., DeVore, 1977, and Cosma et al., 2007). 
(e.g., \citealp{DeVore:1977}, and \citealp{CSS:2007}). In order to choose $J$, we first implement program
(GD) for $J=2$, which we then augment sequentially adding one pair
of cosine and sine basis at a time, up to a representation of order
$J=8$. At each step, we compute a Schwarz Information Criterion \citep{Schwarz:1978}
applied to the primal generalized dual regression problem, exploiting
the strong duality result of Theorem \ref{thm:Formal duality-GDR}
in order to compute its value as $y^{\textrm{T}}e^{*}+\{2+J(K-1)\}\log n$.
Our procedure selects the location-scale representation $J=2$. In
the Supplementary Material, we describe the procedure and report results
from the augmented specifications, which show that our results are
robust to the number of terms included. In order to test for the validity
of the selected model, a complementary procedure that should be explored
in future research is to test for independence of dual regression
solutions and explanatory variables. The test for multivariate independence
proposed by \citet{GQR:2007} constitutes an interesting starting
point for such a development.

All computational procedures can be implemented in the software R \citep{R:2017} 
using open source software packages for nonlinear optimization such as Ipopt or Nlopt, 
and their R interface Ipoptr and Nloptr developed by Jelmer Ypma. Quantile regression 
procedures in the package quantreg have been used to carry
our comparisons.

\begin{figure}
\begin{centering}
\includegraphics[scale=0.45]{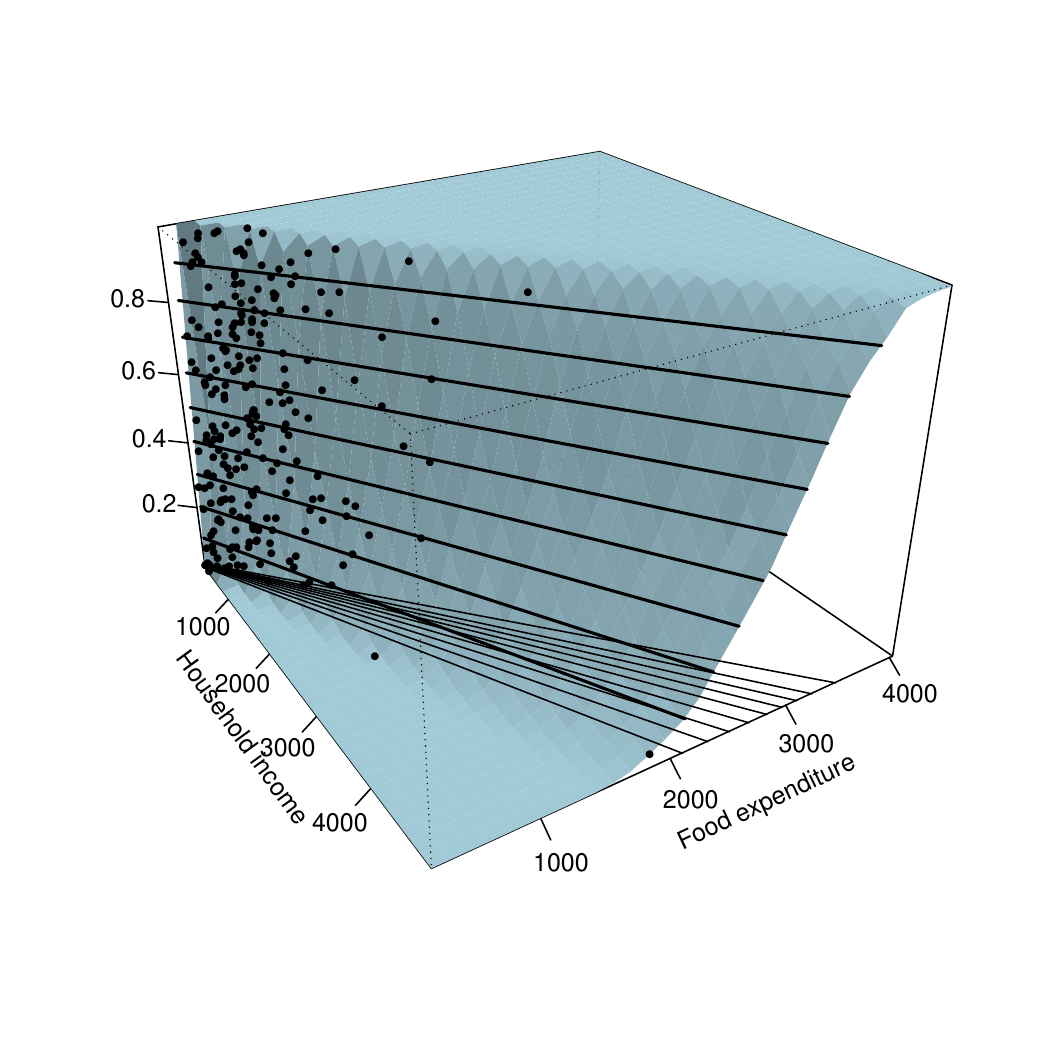} 
\par\end{centering}
\caption{Dual regression estimate of the distribution of food expenditure conditional
on income. Level sets (solid lines) are plotted for a grid of values
ranging from $0\cdot1$ to $0\cdot9$. The projected shadow level
sets yield the respective conditional quantile functions appearing
on the $xy$-plane.}
\label{fig:The-distribution-function} 
\end{figure}

Figure \ref{fig:The-distribution-function} illustrates our results
and plots the estimated distribution of food expenditure conditional
on household income. Estimates $\{u_{i}^{*}\}_{i=1}^{n}$, where $u_{i}^{*}=F_{n}(e_{i}^{*})$,
are used in order to plot each observation in the $xyu$-space with
predicted coordinates $(x_{i},y_{i},u_{i}^{*})$, and the solid lines
give the $u$-level sets for a grid of values $\{0\cdot1,\ldots,0\cdot9\}$.
Although nonstandard, this representation relates to standard quantile
regression plots since the levels of the distribution function give
the conditional quantiles of food expenditure for each value of income.
These are the plotted shadow solid lines corresponding for each
$u$ to dual regression estimates of conditional quantile functions
of food expenditure given household income.
\begin{figure}
\subfloat[]{\includegraphics[scale=0.37]{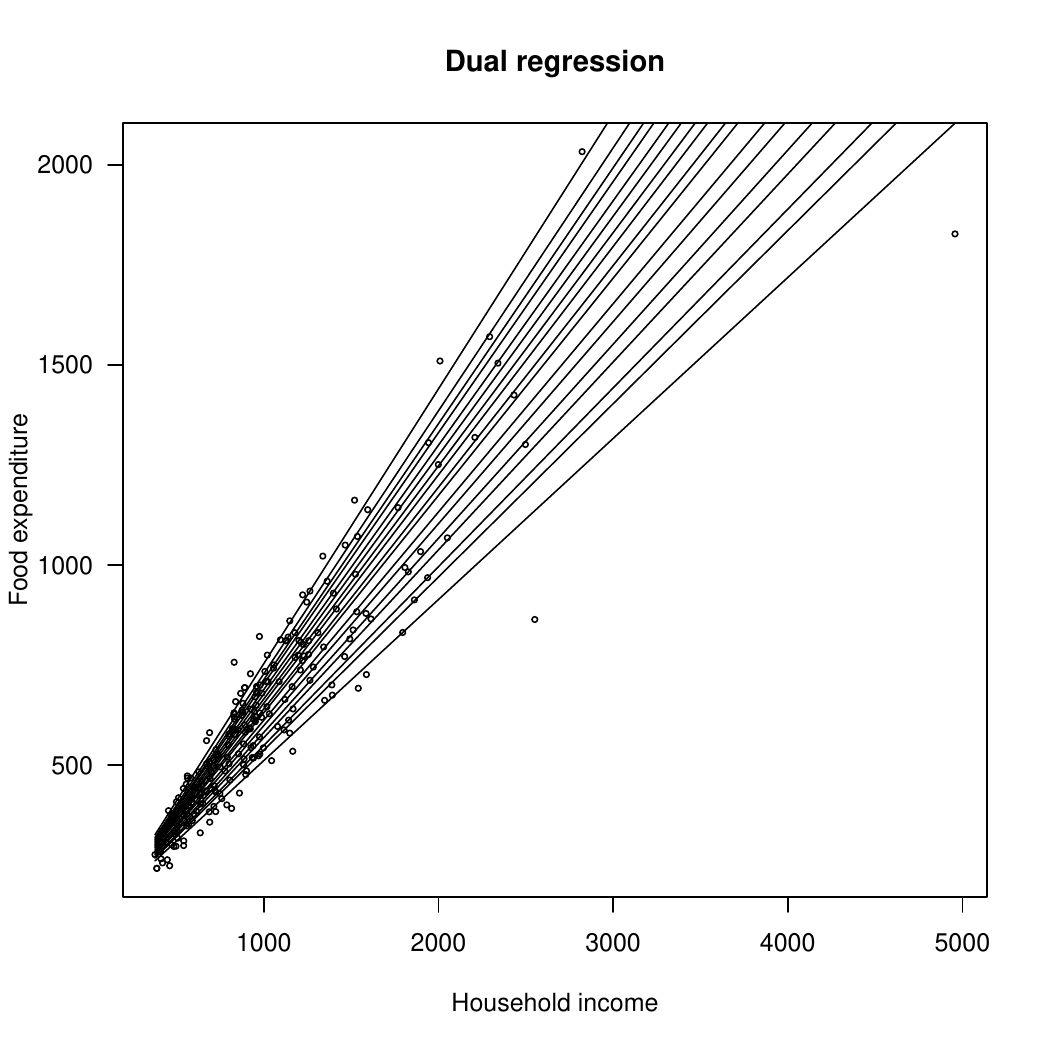}

}\hfill{} \subfloat[]{\includegraphics[scale=0.37]{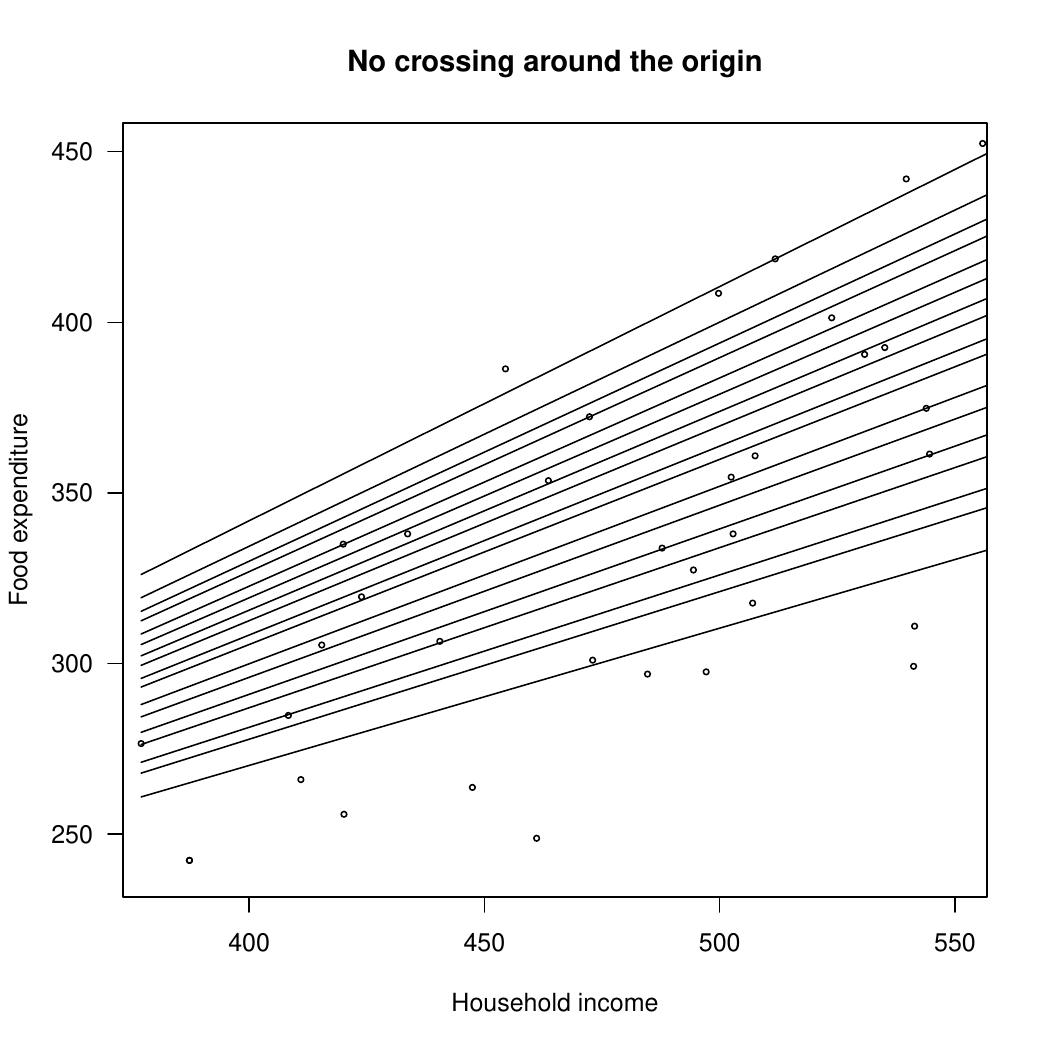}

}\hfill{}

\subfloat[]{\includegraphics[scale=0.37]{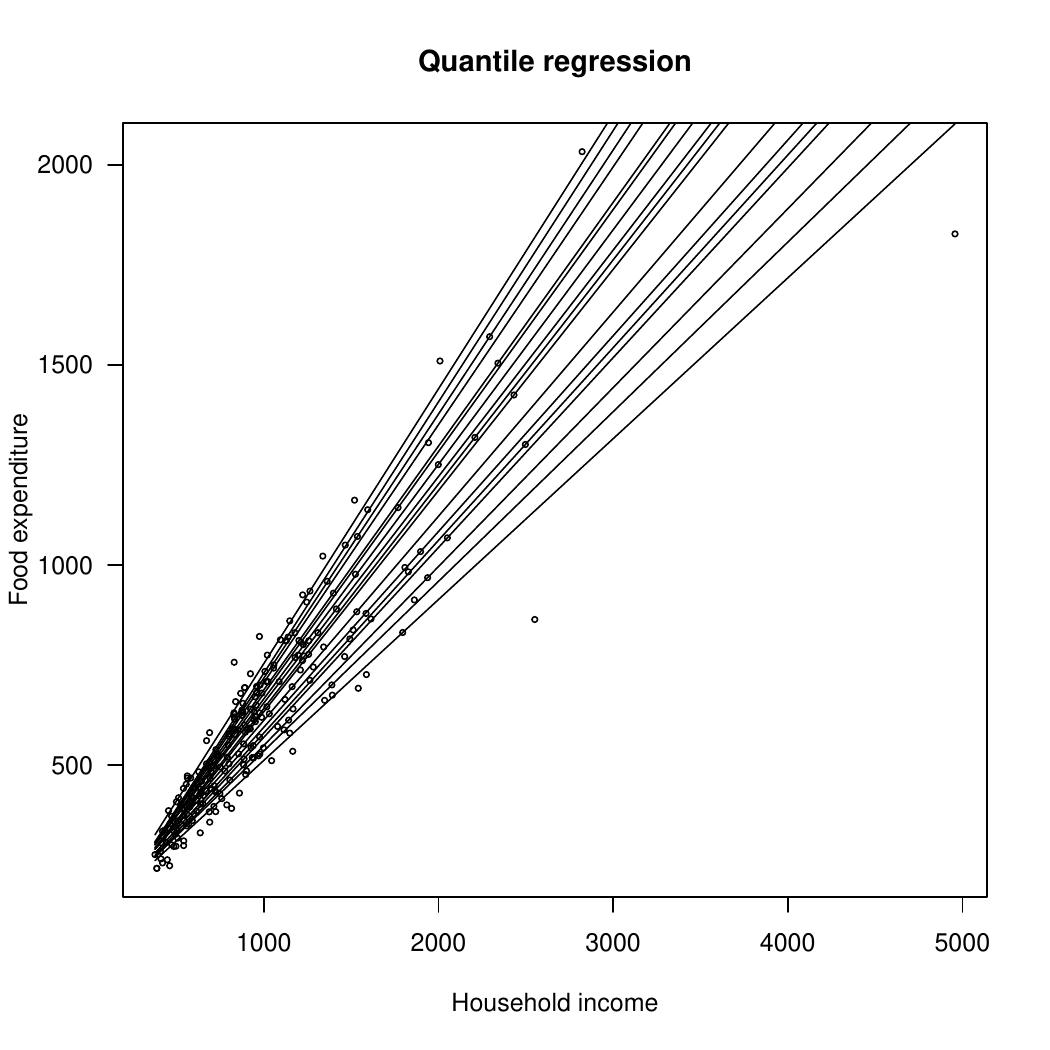}

}\hfill{} \subfloat[]{\includegraphics[scale=0.37]{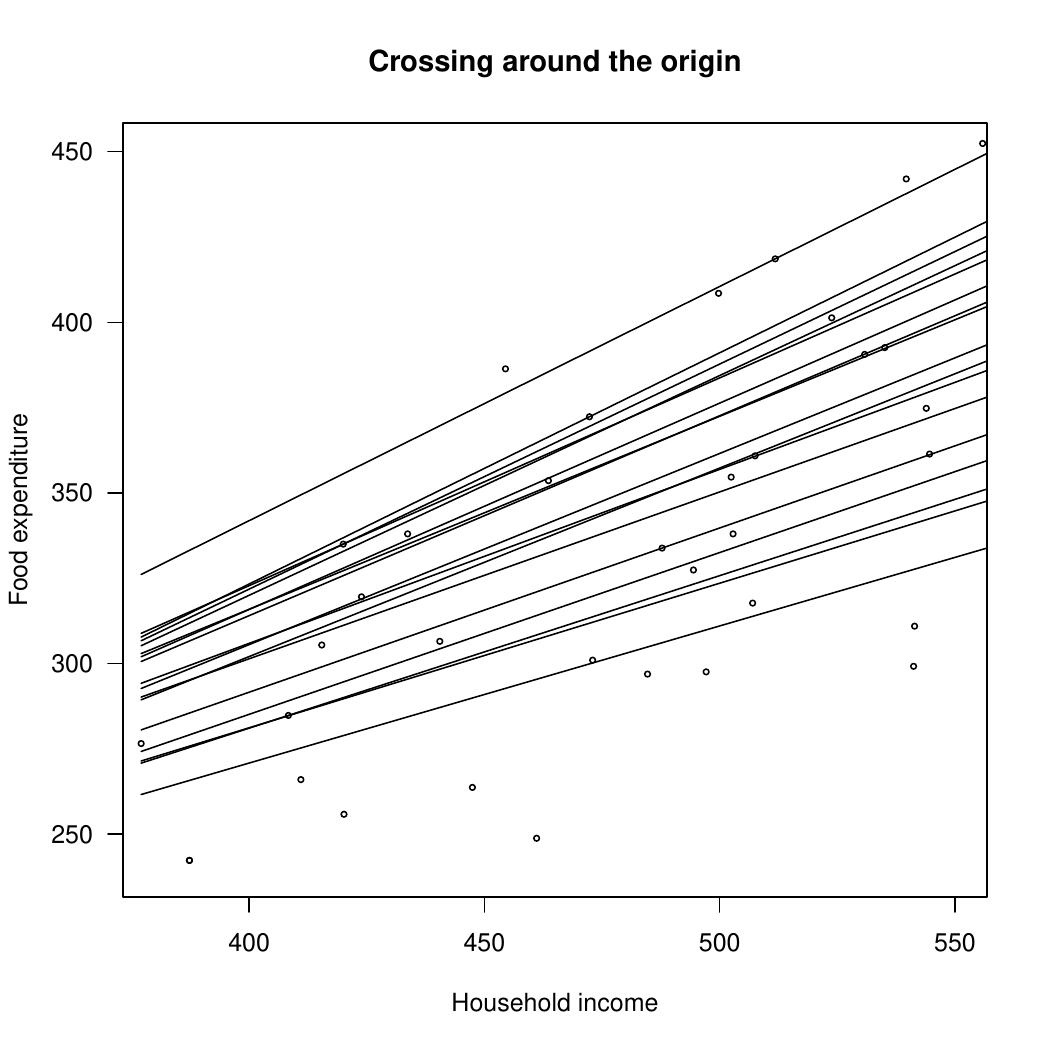}

}\hfill{}

\caption{Scatterplots and dual (a) and quantile (c) regression estimates of
the conditional $\{0\cdot1,0\cdot15,\ldots,0\cdot9\}$ quantile functions
(solid lines) for Engel's data, and their rescaled counterparts ((b),(d)).\label{fig:Scatterplots-and-Dual-1}}
\end{figure}

Figure \ref{fig:The-distribution-function} shows that
the predicted conditional distribution function obtained by dual regression
is indeed endowed with all desired properties. Of particular interest
is the fact that the estimated function is monotone in food expenditure. Also, our estimates satisfy
some basic smoothness requirements across probability levels, in the
food expenditure values. This feature does not typically characterize
estimates of the conditional quantile process by quantile regression
methods, as conditional quantile functions are then estimated sequentially
and independently of each other. The decreasing slope of the distribution
function across values of income provides evidence that the data indeed
follow a heteroscedastic generating process. This is the distributional
counterpart of quantile functions having increasing slope across probability
levels, a feature characterizing the conditional quantile functions
on the $xy$ plane and signalling increasing dispersion in food expenditure
across household income values.

\begin{figure}
\noindent \begin{centering}
\subfloat[]{\includegraphics[scale=0.37]{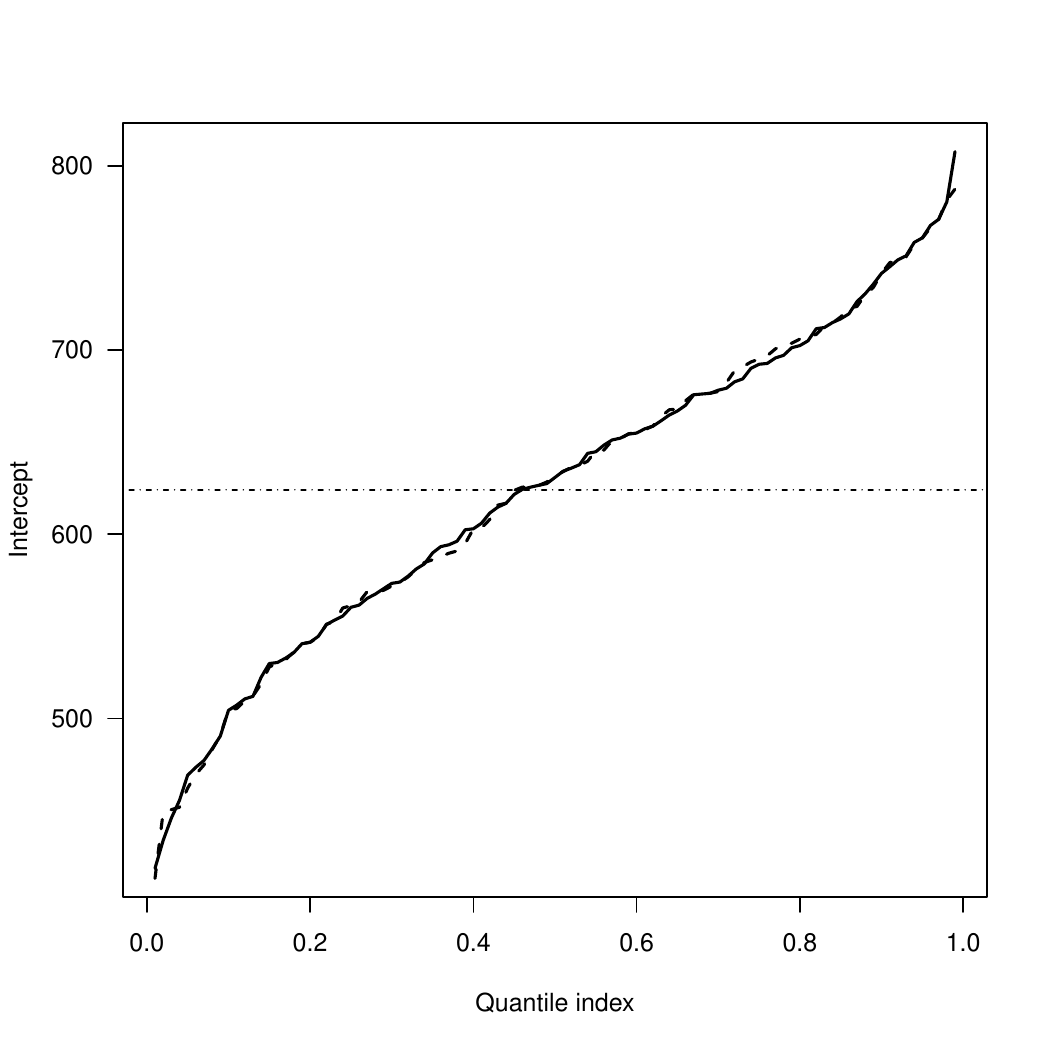}

}\hfill{}\subfloat[]{\includegraphics[scale=0.37]{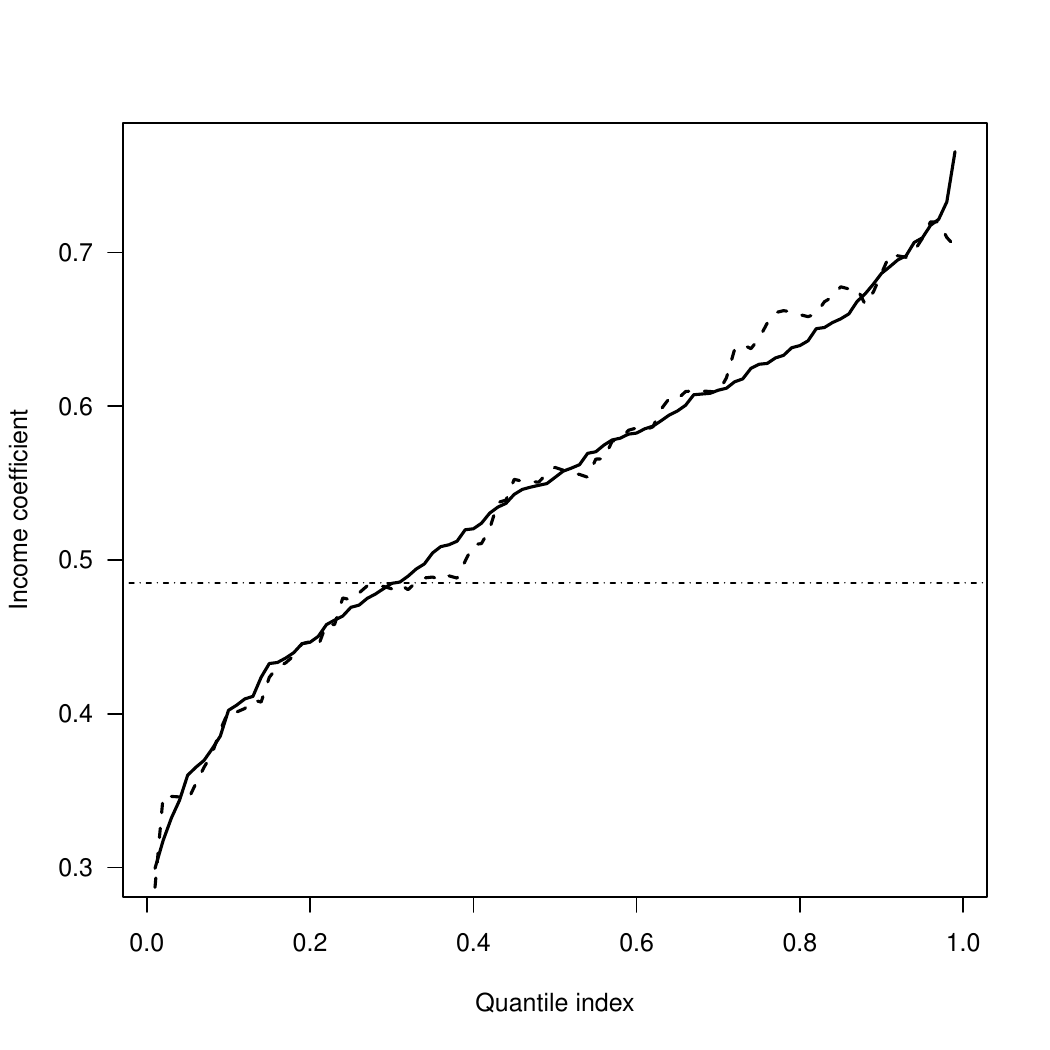}

}\hfill{} 
\par\end{centering}
\caption{Engel coefficient plots revisited. Dual (solid) and quantile (dashes)
regression estimates of the intercept (a) and income (b) coefficients
as a function of the quantile index. Least squares estimates are also
shown (dot-dash). \label{fig:Engel-Coefficient-Plots}}
\end{figure}

Figure \ref{fig:Scatterplots-and-Dual-1} gives the more familiar
quantile regression plots. The plots presented show scatterplots of
Engel's data as well as conditional quantile functions obtained by
dual and quantile regression methods. The rescaled plots in the right
panels of Fig. \ref{fig:Scatterplots-and-Dual-1} highlight some features
of the two procedures. The fitted lines obtained from dual regression
are not subject to crossing in this example, whereas several of the
fitted quantile regression lines actually cross for small values of
household income. Last, the more evenly spread dual regression conditional
quantile functions illustrate the effect of specifying a functional
form for the quantile regression coefficients, while preserving asymmetry
in the conditional distribution of food expenditure.

Figure \ref{fig:Engel-Coefficient-Plots} compares our estimates of
intercept and income coefficients in quantile regression form, with
estimates obtained by quantile regression. For interpretational purposes,
we follow \citet{Koenker:2005} and estimate the functional coefficients
after having recentered household income. This avoids having to interpret
the intercept as food expenditure for households with zero income.
After centering, the intercept coefficient can be interpreted as the
$u$-th quantile of food expenditure for households with mean income. 
Fig. \ref{fig:Engel-Coefficient-Plots}
shows the estimated quantile regression coefficients as a function
of $u$. It illustrates the fact that the flexible structure imposed
by dual regression yields estimates that are indeed smoother than
their quantile regression counterpart, the latter having a somewhat
erratic behaviour around our estimates.

\subsection{Simulations}\label{subsec:simulations}

We give a brief summary of the results of a Monte Carlo simulation
in order to assess the finite-sample properties dual regression. The data-generating process is 
\begin{equation}
y_{i}  =  \alpha_{1}+\beta_{1}\widetilde{x}_{i}+(\alpha_{2}+\beta_{2}\widetilde{x}_{i})\varepsilon_{i},\quad\varepsilon_{i}\sim N(0,1),\label{eq:DGP-1}
\end{equation}
with parameter values calibrated to the empirical application, from
which $4999$ samples are simulated. As a benchmark, we compare generalized dual
regression estimates of the values $F_{Y\mid X}(y_{i}\mid x_{i})$ $(i=1,\ldots,n)$,
to those obtained by applying the inversion procedure of \citet{CFG:2010}
to the quantile regression process. For each simulation, the estimation
and selection procedures are identical to those implemented in the
empirical application.

Table \ref{tab:1} reports a first set of results regarding the accuracy
of conditional distribution function estimates. We report average
estimation errors across simulations of dual regression and quantile
regression estimators, respectively, and their ratio in percentage
terms. Estimation errors are measured in $L^{p}$ norms $\left\Vert \cdot\right\Vert _{p}$, for 
$p=1,2$, and $\infty$, where for $f:\mathbb{R}\mapsto[0,1]$, $\left\Vert f\right\Vert _{p}=\left\{ \int_{\mathbb{R}}\left|f(s)\right|^{p}ds\right\} ^{1/p}$,
and are computed with $e^{*}$ the solutions
to the selected generalized dual regression program. Correct model
selection ranges from $75\%$ of the simulations for $n=100$ to $90\%$
for $n=1000$, providing encouraging evidence about the validity of
the proposed criterion. The results show that for this setup our estimates systematically outperform quantile regression-based
estimates, with the spread in performance increasing with sample
size. Whereas the reduction in average estimation error is between
$8\%$ and $17\%$, depending on the norm, for $n=235$, estimation
error is reduced up to $30\%$ when $n=1000$. The larger reduction
in average errors in $L^{\infty}$ norm reflects the higher accuracy
in estimation of extreme parts of the distribution.

\begin{table}
\def~{\hphantom{0}}{%
\begin{tabular}{lcccccc}
%\\
 Sample size & $L_{GDR}^{1}$ & $L_{GDR}^{1}/L_{QR}^{1}$ & $L_{GDR}^{2}$ & $L_{GDR}^{2}/L_{QR}^{2}$ &  $L_{GDR}^{\infty}$ &  $L_{GDR}^{\infty}/L_{QR}^{\infty}$ \\[5pt]
$n=100$  & $4\cdot11$ & $93\cdot34$ & $5\cdot63$ & $92\cdot59$ & $21\cdot89$ & $89\cdot89$ \\
$n=235$  & $2\cdot70$ & $91\cdot64$ & $3\cdot73$ & $89\cdot73$ & $17\cdot15$ & $82\cdot27$ \\
$n=500$  & $1\cdot85$ & $90\cdot68$ & $2\cdot56$ & $87\cdot98$ & $12\cdot97$ & $75\cdot03$ \\
$n=1000$  & $1\cdot31$ & $90\cdot18$ & $1\cdot83$ & $87\cdot03$ & $9\cdot94$ & $70\cdot12$
\end{tabular}}
\medskip
\caption{$L^{p}$ estimation errors $(\times100)$ %and ratios of $L^{p}$ estimation errors $(\times100)$ 
of generalized dual ($L_{GDR}^{p}$)
and quantile regression ($L_{QR}^{p}$) estimates of $\{F_{Y\mid X}(y_{i}\mid x_{i})\}^{n}_{i=1}$, and their ratios $(\times100)$, for $p=1,2$ and $\infty$.}
\label{tab:1}
\end{table}

In the Supplementary Material, we describe the experiment in detail,
and report results on estimation of quantile regression coefficients
and the distribution of selected models across simulations. We also
include additional simulations that illustrate the empirical performance
of dual regression with multiple covariates and show that it performs
well relative to the noncrossing quantile regression method proposed
by \citet{BRW:2010}.

\section{Discussion}\label{sec:Discussion}

If we designate problems such as (D) and (GD) as already dual,
then their solutions reveal a corresponding primal. Typically, the
Lagrange multipliers of the dual appear as parameters in the primal,
and the primal has an interpretation as a data-generating process.
So perhaps not surprisingly the constraints on the construction of
the stochastic elements have shadow values that are parameters of
a data-generating representation. In this way the relation between
identification and estimation is made perspicuous: a parameter
of the data-generating process is the Lagrange multiplier of a specific
constraint on the construction of the stochastic element, so to specify
that some parameters are non-zero and others are zero is to say that
some constraints are in the large-sample limit binding and others
are not.

Another way of expressing this is to say that when a primal corresponds
to the data-generating process, additional moment conditions are superfluous:
they will in the limit attract Lagrange multiplier values of zero
and consequently not affect the value of the program %(the objective
%function)
nor the solution. In a sense, this is obvious: the parameters
of the primal can typically be identified and estimated through an
$M$--estimation problem that will generate $K$ equations to be solved
for the $K$ unknown parameters. Nonetheless, the recognition that
the only moment conditions that contribute to enforcing the independence
requirement are those whose imposition simultaneously reduces the
objective function while providing multipliers that are coefficients
in the stochastic representation of $Y$ suggests the futility of
portmanteau approaches (e.g., those based on characteristic functions)
to imposing independence. The dual formulation reveals that to specify
the binding moment conditions is to specify an approximating
data-generating process representation, which then can be extrapolated
to provide estimates of objects of interest beyond the $n$ explicitly
estimated values of $\varepsilon_{i}$ that characterize the sample
and the definition of the mathematical program.

As is well understood in mathematical programming, dual solutions
provide lower bounds on the values obtained by primal problems. In
the generic form of the problems we have considered here there is
no gap between the primal and dual values; hence in econometrics these
problems are said to display point identification. We conjecture
that the problems without point identification do have gaps between
their dual and primal values, and that this characterization will
enhance our understanding.

\appendix

\section{Proof of Theorem 2}

For $\Lambda\in\mathbb{R}$, the Lagrangian of the infeasible problem
(IGD) is
\[
\mathscr{L}^{IGD}(e,\Lambda)=y^{\textrm{T}}e-\Lambda\left\{\sum_{i=1}^{n}\widetilde{H}_{x_{i}}(e_{i})-S_{n}\right\}\quad(e\in\mathbb{R}^{n}).
\]
By definition of $\widetilde{H}_{x_{i}}$ and continuous differentiability
of $H_{x_{i}}$, for each $x_{i}$, the Fundamental Theorem of Calculus
implies the $n$ first-order conditions 
\begin{equation}
\nabla_{e_{i}}\mathscr{L}^{IGD}=y_{i}-\Lambda H_{x_{i}}(e_{i})=0,\qquad(i=1,\ldots,n),\label{eq:FOCs}
\end{equation}
The $n$ second-order conditions 
\begin{equation}
\nabla_{e_{i}e_{i}}\mathscr{L}^{IGD}=-\Lambda H_{x_{i}}'(e_{i})<0,\qquad(i=1,\ldots,n),\label{eq:SOCs}
\end{equation}
are satisfied if and only if $\Lambda>0$, since $H'_{x_{i}}$ is
strictly positive for each $x_{i}$. Since $y=0$ cannot hold under
Condition \ref{Data}, with probability one, (\ref{eq:FOCs}) rules
out $\Lambda=0$. Moreover, for $\Lambda<0$, (\ref{eq:SOCs}) implies
that the map $e\mapsto\mathscr{L}^{IGD}(e,\Lambda)$ is strictly
convex over the real line, since $\inf_{e\in\mathbb{R}}H'_{x_{i}}(e)\geq\tau>0$
for each $x_{i}$, and hence unbounded above. Therefore we only need
to show that the pair $(1,\varepsilon_{o})$ is the unique pair in
$(0,\infty)\times\mathbb{R}^{n}$ that satisfies (\ref{eq:FOCs}).

By strict monotonicity of $H_{x_{i}}$, the inverse function $H_{x_{i}}^{-1}$
is well-defined, for each $x_{i}$, and a solution to (\ref{eq:FOCs})
is $e_{i}=H_{x_{i}}^{-1}(y_{i}/\Lambda)$ $(i=1,\ldots,n)$. Substituting
into the constraint of (IGD) yields 
\begin{equation}
\sum_{i=1}^{n}\widetilde{H}_{x_{i}}\left\{ H_{x_{i}}^{-1}\left(\frac{y_{i}}{\Lambda}\right)\right\} -S_{n}=0.\label{eq:Constraint}
\end{equation}
By Lemma \ref{lem:Uniqueness} below, $\Lambda=1$ is the unique solution
to (\ref{eq:Constraint}) such that $\Lambda>0$. Since $H_{x_{i}}^{-1}(y_{i}/\Lambda)=\varepsilon_{i}$ 
 $(i=1,\ldots,n)$ for $\Lambda=1$, strict concavity of $\mathscr{L}^{IGD}$
for $\varLambda>0$ implied by (\ref{eq:SOCs}) shows that $(1,\varepsilon_{o})$
is the unique pair in $(0,\infty)\times\mathbb{R}^{n}$ that satisfies
(\ref{eq:FOCs}).

%\begin{lemmaS1*}
\begin{lemma}
Under the conditions of Theorem 2, $\Lambda=1$ is the
unique solution to the equation 
\begin{equation}
\sum_{i=1}^{n}\widetilde{H}_{x_{i}}\left\{ H_{x_{i}}^{-1}\left(\frac{y_{i}}{\Lambda}\right)\right\} -S_{n}=0\label{eq:Constraint-1}
\end{equation}
such that $\Lambda>0$.
\label{lem:Uniqueness}
\end{lemma}

\begin{proof}
We first show that Equation (\ref{eq:Constraint-1})
is the first-order condition of the infeasible generalized dual regression
primal problem $\min_{\Lambda>0}Q_{n}^{IGD}(\Lambda)$, where 
\[
Q_{n}^{IGD}(\Lambda)=\sum_{i=1}^{n}y_{i}H_{x_{i}}^{-1}\left(\frac{y_{i}}{\Lambda}\right)-\Lambda\left[\sum_{i=1}^{n}\widetilde{H}_{x_{i}}\left\{ H_{x_{i}}^{-1}\left(\frac{y_{i}}{\Lambda}\right)\right\} -S_{n}\right]\quad(\Lambda>0),
\]
and then show that $Q_{n}^{IGD}(\Lambda)$ admits $\Lambda=1$ as
its unique minimum.

Step 1. Define the Lagrange dual function (\citealp{BV:2004},
Chapter 5) $Q_{n}^{IGD}(\varLambda)\equiv\sup_{e\in\mathbb{R}^{n}}\mathscr{L}^{IGD}(e,\varLambda)$,
for $\varLambda>0$. In order to derive $Q_{n}^{IGD}(\varLambda)$,
we show that the maximum of the map $e\mapsto\mathscr{L}^{IGD}(e,\varLambda)$
is attained and is unique, and evaluate $e\mapsto\mathscr{L}^{IGD}(e,\varLambda)$
at this value.

For $\varLambda>0$ and $c\in\mathbb{R}$, consider the level sets
$\mathcal{B}_{c}(\varLambda)=\{e\in\mathbb{R}^{n}:-\mathscr{L}^{IGD}(e,\varLambda)\leq c\}$
of $-\mathscr{L}^{IGD}$. These sets are compact. Given $e_{1},e_{2}\in\mathcal{B}_{c}(\varLambda)$,
let $t=||e_{1}-e_{2}||$ and $u=\frac{e_{1}-e_{2}}{||e_{1}-e_{2}||}$,
so that $||u||=1$ and $e_{1}=e_{2}+tu$. Thus, by definition of $e_{1}$,
a second-order Taylor expansion of $t\mapsto-\mathscr{L}^{IGD}(e_{2}+tu,\varLambda)$
around $t=0$ yields, for some $\bar{e}$ on the line connecting $e_{1}$
and $e_{2}$,
\begin{align*}
c\geq-\mathscr{L}^{IGD}(e_{1},\varLambda) & =-\mathscr{L}^{IGD}(e_{2}+tu,\varLambda)\\
 & =-\mathscr{L}^{IGD}(e_{2},\varLambda)-t\nabla_{e}\mathscr{L}^{IGD}(e_{2},\varLambda)^{T}u-\frac{t^{2}}{2}u^{T}\nabla_{ee}\mathscr{L}^{IGD}(\bar{e},\varLambda)u\\
 & \geq-\mathscr{L}^{IGD}(e_{2},\varLambda)-t||\nabla_{e}\mathscr{L}^{IGD}(e_{2},\varLambda)^{T}||+\Lambda\tau\frac{t^{2}}{2},
\end{align*}
where the last inequality follows from $-\nabla_{ee}\mathscr{L}^{IGD}(\bar{e},\varLambda)=\Lambda\textrm{diag}\{H_{x_{i}}'(\bar{e}_{i})\}$
and the uniform lower bound on $H_{x_{i}}'$ for each $x_{i}$ which
implies that $-\nabla_{ee}\mathscr{L}^{IGD}(\bar{e},\varLambda)$
is positive definite. For $e_{2}\in\mathcal{B}_{c}(\varLambda)$,
the above inequality implies that $t$ is bounded and therefore $\mathcal{B}_{c}(\varLambda)$
is bounded. Since $e\mapsto-\mathscr{L}(e,\varLambda)$ is continuous
over $\mathbb{R}^{n}$, $\mathcal{B}_{c}(\varLambda)$ is also closed.
It then follows from the Weierstrass theorem that there exists $e(\varLambda)\in\arg\min_{e\in\mathbb{R}^{n}}\{-\mathscr{L}^{IGD}(e,\varLambda)\}=\arg\max_{e\in\mathbb{R}^{n}}\mathscr{L}^{IGD}(e,\varLambda)$.

Since the Hessian matrix of the map $e\mapsto\mathscr{L}^{IGD}(e,\varLambda)$
is negative definite for all $\varLambda>0$, $e\mapsto\mathscr{L}^{IGD}(e,\varLambda)$
is strictly concave with unique maximum $e(\varLambda)$, for all
$\varLambda>0$. Upon using first-order conditions (\ref{eq:FOCs}),
direct substitution yields $\mathscr{L}^{IGD}\{e(\varLambda),\Lambda\}=Q_{n}^{IGD}(\Lambda)$,
the maximum of the map $e\mapsto\mathscr{L}^{IGD}(e,\Lambda)$, for
all $\Lambda>0$.

Step 2. The function $Q_{n}^{IGD}(\Lambda)$ is strictly convex for
$\varLambda>0$: since $H_{x_{i}}$ is continuously differentiable
for each $x_{i}$ by assumption, by the inverse function theorem $H_{x_{i}}^{-1}$
is continuously differentiable for each $x_{i}$, and there are the
following derivatives: 
\begin{align}
\nabla_{\Lambda}H_{x_{i}}^{-1}\left(\frac{y_{i}}{\Lambda}\right) & =-\frac{1}{H_{x_{i}}'\left\{ H_{x_{i}}^{-1}\left(\frac{y_{i}}{\Lambda}\right)\right\} }\frac{y_{i}}{\Lambda^{2}}\label{eq:Deriv1}\\
\nabla_{\Lambda}\widetilde{H}_{x_{i}}\left\{ H_{x_{i}}^{-1}\left(\frac{y_{i}}{\Lambda}\right)\right\}  & =-\frac{y_{i}}{\Lambda}\frac{1}{H_{x_{i}}'\left\{ H_{x_{i}}^{-1}\left(\frac{y_{i}}{\Lambda}\right)\right\} }\frac{y_{i}}{\Lambda^{2}},\label{eq:Deriv2}
\end{align}
for every $x_{i}$, $y_{i}$ and $\varLambda>0$. Upon using (\ref{eq:Deriv1})
and (\ref{eq:Deriv2}), $Q_{n}^{IGD}(\Lambda)$ has first-order conditions
(\ref{eq:Constraint-1}), and the second-order conditions 
\[
\nabla_{\Lambda\Lambda}Q_{n}^{IGD}=\frac{1}{\Lambda}\sum_{i=1}^{n}\frac{1}{H_{x_{i}}'\left\{ H_{x_{i}}^{-1}\left(\frac{y_{i}}{\Lambda}\right)\right\} }\left(\frac{y_{i}}{\Lambda}\right)^{2}>0
\]
are satisfied for all $\varLambda>0$ since $H'_{x_{i}}>0$ for each
$x_{i}$. Therefore, $Q_{n}^{IGD}(\Lambda)$ is strictly convex for
all $\varLambda>0$ and admits at most one minimum. Since $H_{x_{i}}^{-1}(y_{i}/\Lambda)=\varepsilon_{oi}$
 $(i=1,\ldots,n)$ for $\Lambda=1$, and $S_{n}=\sum_{i=1}^{n}\widetilde{H}_{x_{i}}(\varepsilon_{oi})$
by definition, $\Lambda=1$ is also feasible. The result follows.\end{proof}

\section{Proofs of Theorems 1 and 3}

\subsection{Proof of Theorem 1}

Theorem 1 is a corollary of Theorem 3,
upon substituting $x_{i}$ to $x_{i}^{c}$ and setting $J=2$.

\subsection{Preliminary lemmas}
We establish the equivalence (IGD)--(GD) and convexity of (GP).

%\begin{lemmaS2*}
\begin{lemma}
If Conditions \ref{Data}, \ref{Design}, \ref{FR} and \ref{ExistenceGDR} hold with %$\Omega_{n}=\textrm{diag}[H_{x_{i}}'\{e(y_{i},x_{i},\theta);\theta\}]$,
$\omega=\phi(\theta)$,
for all $\theta\in\Theta_{n}$, then the infeasible problem (IGD) admits the equivalent formulation (GD).
\label{lem:equivalenceIGD-GD}
\end{lemma}
%\end{lemmaS2}
\begin{proof}
Letting $\widetilde{H}_{x_{i}}(e_{oi};\theta)=\int_{0}^{e_{oi}}H_{x_{i}}(s;\theta)ds$,
the corresponding expression is 
\begin{equation}
\widetilde{H}_{x_{i}}(e_{oi};\theta)=\sum_{j=1}^{J}(\theta_{j}^{\textrm{T}} x_{ij}^{c})\widetilde{h}_{j}(e_{oi})\quad(e_{oi}\in\mathbb{R}).\label{eq:Htilde}
\end{equation}
Given the form of $\widetilde{H}_{x_{i}}(\cdot;\theta)$, the constraint
$\sum_{i=1}^{n}\widetilde{H}_{x_{i}}(e_{i};\theta)=S_{n}$ in (IGD)
can be simplified using
\[
S_{n}=\sum_{i=1}^{n}\widetilde{H}_{x_{i}}(e_{oi};\theta)=\sum_{i=1}^{n}\sum_{j=1}^{J}(\theta_{j}^{\textrm{T}} x_{ij}^{c})\widetilde{h}_{j}(e_{oi})=\sum_{j=1}^{J}\theta_{j}^{\textrm{T}} \left\{ \sum_{i=1}^{n}x_{ij}^{c}\widetilde{h}_{j}(e_{oi})\right\} =0,
\]
by definition of $S_{n}$ in Theorem 2, expansion (\ref{eq:Htilde}),
and the properties of $e_{oi}$ assumed in Condition \ref{ExistenceGDR}.
Therefore, the infeasible problem (IGD) becomes 
\[
\max_{e\in\mathbb{R}^{n}}\left\{ y^{\textrm{T}}e : \sum_{j=1}^{J}\sum_{i=1}^{n}(\theta_{j}^{\textrm{T}} x_{ij}^{c})\widetilde{h}_{j}(e_{i})=0\right\} ,
\]
with Lagrangian
\[
\mathscr{L}(e,\varLambda)=y^{\textrm{T}}e-\Lambda\left\{ \sum_{i=1}^{n}\widetilde{H}_{x_{i}}(e_{i};\theta)-S_{n}\right\} =y^{\textrm{T}}e-\Lambda\sum_{j=1}^{J}\sum_{i=1}^{n}(\theta_{j}^{\textrm{T}} x_{ij}^{c})\widetilde{h}_{j}(e_{i})\quad(e\in\mathbb{R}^{n}),
\]
for $\varLambda\in\mathbb{R}$. For all $\theta\in\Theta_{n}$, the
map $e_{i}\mapsto H_{x_{i}}(e_{i};\theta)$ satisfies the conditions
of Theorem 2, which implies that $\Lambda=1$ and $e=e_{o}$,
by application of Theorem 2 upon substituting $H_{x_{i}}(\cdot;\theta)$
for $H_{x_{i}}(\cdot)$ and $e_{oi}$ for $\varepsilon_{i}$ $(i=1,\ldots,n)$. 

Adding $\theta$ to the choice variables of the optimization problem,
we obtain the $\textrm{dim}(\theta)$ additional constraints 
\begin{eqnarray}
\nabla_{\theta_{j}}\mathscr{L}=-\sum_{i=1}^{n}x_{ij}^{c}\widetilde{h}_{j}(e_{i}) & = & 0\quad(j=1,\ldots,J).\label{eq:21n}
\end{eqnarray}
Equation (\ref{eq:21n}) can be directly appended to the objective
$\max_{e\in\mathbb{R}^{n}}\;y^{\textrm{T}}e$ to obtain the optimization
problem (GD) in which the Lagrange multiplier is $\theta$. By part
(iii) of Theorem 3, problem (GD) admits
a unique optimal solution $e^{*}$ over $\mathbb{R}^{n}$. Since $e_{o}$
is a feasible solution by Condition \ref{ExistenceGDR}, $e^{*}=e_{o}$.
It follows that (GD) and (IGD) are equivalent.
\end{proof}

%\begin{lemmaS3*}
\begin{lemma}
If Conditions \ref{Data}, \ref{Design} and \ref{FR} hold with %$\Omega_{n}=\textrm{diag}[H_{x_{i}}'\{e(y_{i},x_{i},\theta);\theta\}]$, 
$\omega=\phi(\theta)$, 
for all $\theta\in\Theta_{n}$, 
then, the first-order conditions of (GP) are $\sum_{i=1}^{n}x_{ij}^{c}\widetilde{h}_{j}\{e(y_{i},x_{i},\theta)\}=0$ 
$(j=1,\ldots,J)$, and the Hessian matrix of the objective function
of (GP) is positive definite for all $\theta\in\Theta_{n}$.
\label{lem:convexity}
\end{lemma}
%\end{lemmaS3}
\begin{proof}For $\theta\in\Theta_{n}$, define $Q_{n}(\theta)=\sum_{i=1}^{n}L(x_{i},y_{i},\theta)$,
with $L(x_{i},y_{i},\theta)$ defined as
%as
\begin{equation}
L(x_{i},y_{i},\theta)=\sum_{j=2}^{J}(\theta_{j}^{\textrm{T}} x_{ij}^{c})\left[h_{j}\{e(y_{i},x_{i},\theta)\}e(y_{i},x_{i},\theta)-\widetilde{h}_{j}\{e(y_{i},x_{i},\theta)\}\right],\label{eq:L(X,Y)}
\end{equation}
%given in (\ref{eq:L(X,Y)}), 
and 
let $e_{i}=e(y_{i},x_{i},\theta)$,
$\eta_{j}(e_{i})=h_{j}(e_{i})e_{i}-\widetilde{h}_{j}(e_{i})$, $(i=1,\dots,n; j=1,\ldots,J)$.
For $j=1,\ldots,J$ and $\theta\in\Theta_{n}$, the derivative of
$Q_{n}$ with respect to $\theta_{j}$ satisfies
\[
\nabla_{\theta_{j}}Q_{n}(\theta)=\sum_{i=1}^{n}\sum_{l=2}^{J}(\theta_{l}^{\textrm{T}} x_{il}^{c})\eta_{l}'(e_{i})\nabla_{\theta_{j}}e_{i}+\sum_{i=1}^{n}x_{ij}^{c}\eta_{j}(e_{i})=\sum_{i=1}^{n}H_{x_{i}}'(e_{i};\theta)e_{i}\nabla_{\theta_{j}}e_{i}+\sum_{i=1}^{n}x_{ij}^{c}\eta_{j}(e_{i}),
\]
upon substituting $\eta_{l}'=h'_{l}(e_{i})e_{i}$ and by definition
of $H_{x_{i}}'(e_{i};\theta)$. Using
\[
\nabla_{\theta_{j}}e_{i}=-\nabla_{\theta_{j}}H_{x_{i}}(e_{i};\theta)\{H_{x_{i}}'(e_{i};\theta)\}^{-1}=-x_{ij}^{c}h_{j}(e_{i})\{H_{x_{i}}'(e_{i};\theta)\}^{-1},
\]
and the definition of $\eta_{j}$, for all $\theta\in\Theta_{n}$
we obtain
\begin{equation}
\nabla_{\theta_{j}}Q_{n}(\theta)=-\sum_{i=1}^{n}x_{ij}^{c}h_{j}(e_{i})e_{i}+\sum_{i=1}^{n}x_{ij}^{c}\{h_{j}(e_{i})e_{i}-\widetilde{h}_{j}(e_{i})\}=-\sum_{i=1}^{n}x_{ij}^{c}\widetilde{h}_{j}(e_{i}),\quad(j=1,\ldots,J).\label{eq:fock}
\end{equation}
Letting $p_{i}=(1,e_{i})^{\textrm{T}}$ and $q_{i}=\{h_{3}(e_{i}),\ldots,h_{J}(e_{i})\}^{\textrm{T}}$, upon using (\ref{eq:fock}) the Hessian matrix is
\[
H_{n}=\sum_{i=1}^{n}\left[\begin{array}{cc}
\frac{x_{i}^{c}x_{i}^{c\textrm{T}}}{H_{x_{i}}'(e_{i};\theta)}\otimes p_{i}p_{i}^{\textrm{T}} & \frac{x_{i}^{c}\widetilde{x}_{i}^{c\textrm{T}}}{H_{x_{i}}'(e_{i};\theta)}\otimes p_{i}q_{i}^{\textrm{T}}\\
\frac{\widetilde{x}_{i}^{c}x_{i}^{c\textrm{T}}}{H_{x_{i}}'(e_{i};\theta)}\otimes q_{i}p_{i}^{\textrm{T}} & \frac{\widetilde{x}_{i}^{c}\widetilde{x}_{i}^{c\textrm{T}}}{H_{x_{i}}'(e_{i};\theta)}\otimes q_{i}q_{i}^{\textrm{T}}
\end{array}\right]\equiv\left[\begin{array}{cc}
H_{11,n} & H_{12,n}\\
H_{21,n} & H_{22,n}
\end{array}\right].
\]
Suppose $H_{11,n}$ is positive definite for all $\theta\in\Theta_{n}$.
Positive definiteness of $H_{11,n}$ then implies that $H_{n}$ is
positive definite for all $\theta\in\Theta_{n}$ if and only if the
Schur complement of $H_{11,n}$ in $H_{n}$ is positive definite (\citealp{BV:2004},
Appendix A.5.5) for all $\theta\in\Theta_{n}$, i.e., if and only if
the determinant of $D_{n}=H_{22,n}-H_{21,n}H_{11,n}^{-1}H_{12,n}$
is strictly positive, for all $\theta\in\Theta_{n}$. Letting $\Xi_{n}=H_{21,n}H_{11,n}^{-1}$
for all $\theta\in\Theta_{n}$, $D_{n}$ is equal to 
\begin{equation}
\sum_{i=1}^{n}\left[\left\{ \frac{\widetilde{x}_{i}^{c}\otimes q_{i}}{\{H_{x_{i}}'(e_{i};\theta)\}^{1/2}}-\Xi_{n}\frac{x_{i}^{c}\otimes p_{i}}{\{H_{x_{i}}'(e_{i};\theta)\}^{1/2}}\right\} \left\{ \frac{\widetilde{x}_{i}^{c}\otimes q_{i}}{\{H_{x_{i}}'(e_{i};\theta)\}^{1/2}}-\Xi_{n}\frac{x_{i}^{c}\otimes p_{i}}{\{H_{x_{i}}'(e_{i};\theta)\}^{1/2}}\right\} ^{\textrm{T}}\right],\label{eq:Hess2}
\end{equation}
a positive semidefinite matrix, and equal to zero if and only if 
\begin{equation}
\widetilde{x}_{i}^{c}\otimes q_{i}=\Xi_{n}(x_{i}^{c}\otimes p_{i})\quad(i=1,\ldots,n);\label{eq:system-1-1}
\end{equation}
this is an application of the Cauchy-Schwarz inequality for matrices
stated in \citet{Tripathi:1999}. Under Condition \ref{FR}, system
(\ref{eq:system-1-1}) cannot hold, with probability 1, for all $\theta\in\Theta_{n}$.

Finally, a similar argument shows that, under Condition \ref{Design},  $H_{11,n}$ is positive definite
for all $\theta\in\Theta_{n}$ if and only if 
\begin{equation}
x_{i}^{c}e_{i}=\Upsilon_{n}x_{i}^{c} \quad (i=1,\ldots,n),\label{eq:system-1-1-1}
\end{equation}
where 
\[
\Upsilon_{n}=\left[\sum_{i=1}^{n}\frac{x^{c}_{i}x^{c \textrm{T}}_{i}}{\{H_{x_{i}}'(e_{i};\theta)\}^{1/2}}e_{i}\right]\left[\sum_{i=1}^{n}\frac{x^{c}_{i}x^{c \textrm{T}}_{i}}{\{H_{x_{i}}'(e_{i};\theta)\}^{1/2}}\right]^{-1}.
\]
In particular, with $\Upsilon_{n,1}$ denoting the first row of $\Upsilon$,
since $x_{i}^{c}$ includes an intercept system (\ref{eq:system-1-1-1})
implies $e_{i}=\Upsilon_{n,1}x_{i}^{c}$ $(i=1,\ldots,n)$, for all $\theta\in\Theta_{n}$, which cannot hold under Condition \ref{Data},
with probability 1.\end{proof}

\subsection{Proof of Theorem 3}

The equivalence result follows by Lemma \ref{lem:equivalenceIGD-GD}. For $\theta\in\mathbb{R}^{2+J(K-1)}$, define the Lagrangian for (GD)
as
\[
\mathscr{L}(e,\theta)=\sum_{i=1}^{n}(y_{i}-\theta_{1}^{\textrm{T}} x_{i}^{c})e_{i}-\frac{1}{2}\sum_{i=1}^{n}(\theta_{2}^{\textrm{T}} x_{i}^{c})(e_{i}^{2}-1)-\sum_{i=1}^{n}\sum_{j=3}^{J}(\theta_{j}^{\textrm{T}} \widetilde{x}_{i}^{c})\widetilde{h}_{j}(e_{i})\quad(e\in\mathbb{R}^{n}),
\]
with $n$ first-order conditions
\begin{equation}
y_{i}=H_{x_{i}}(e_{i};\theta)\quad(i=1,\ldots,n),\label{eq:FOCsGD}
\end{equation}
and denote any vector in $\mathbb{R}^{n}$ satisfying (\ref{eq:FOCsGD})
by $e(\theta)$, and the $i$th element of $e(\theta)$ by $e(y_{i},x_{i},\theta)$.

\textit{Proof of part (i).} Define the Lagrange dual
function (\citealp{BV:2004}, Chapter 5) $Q_{n}(\theta)\equiv\sup_{e\in\mathbb{R}^{n}}\mathscr{L}(e,\theta)$
for $\theta\in\Theta_{n}$. In order to derive $Q_{n}(\theta)$, we
show that the maximum of the mapping $e\mapsto\mathscr{L}(e,\theta)$
is attained and is unique, and evaluate $e\mapsto\mathscr{L}(e,\theta)$
at this value.

Step 1. We show that the map $e\mapsto\mathscr{L}(e,\theta)$ admits
at least one maximum in $\mathbb{R}^{n}$ for all $\theta\in\Theta_{0,n}$.
Since $\Theta_{n}\subseteq\Theta_{0,n}$, existence of a maximum then
holds for all $\theta\in\Theta_{n}$. For $\theta\in\Theta_{0,n}$
and $c\in\mathbb{R}$, consider the level sets $\mathcal{B}_{c}(\theta)=\{e\in\mathbb{R}^{n}:-\mathscr{L}(e,\theta)\leq c\}$
of $-\mathscr{L}$. These sets are compact. Consider a sequence $(e_{(m)})$
in $\mathbb{R}^{n}$ such that $||e_{(m)}||\rightarrow\infty$ as
$m\rightarrow\infty$. Let $z_{(m)}=\frac{e_{(m)}}{||e_{(m)}||}$,
a bounded sequence with unit norm. By the Bolzano-Weierstrass theorem
there exists a convergent subsequence $z_{(m_{l})}$, $m_{l}\rightarrow\infty$
as $l\rightarrow\infty$, with limit $z_{o}$, say. Then, using that
$\theta_{2}^{\textrm{T}} x_{i}>0$  $(i=1,\ldots,n)$, and $\max_{j=3,\ldots,J}||\widetilde{h}_{j}||_{\infty}$
is bounded, for $\theta\in\varTheta_{0,n}$ 
\begin{align*}
-\mathscr{L}(e_{(m_{l})},\theta) & =-||e_{(m_{l})}||\sum_{i=1}^{n}(y_{i}-\theta_{1}^{\textrm{T}} x_{i})z_{i,(m_{l})}+||e_{(m_{l})}||^{2}\sum_{i=1}^{n}\frac{1}{2}(\theta_{2}^{\textrm{T}} x_{i})z_{i,(m_{l})}^{2}\\
 & +\frac{1}{2}\sum_{i=1}^{n}(\theta_{2}^{\textrm{T}} x_{i})+\sum_{i=1}^{n}\sum_{j=3}^{J}(\theta_{j}^{\textrm{T}} \widetilde{x}_{i})\widetilde{h}_{j}(||e_{(m_{l})}||z_{i,(m_{l})})\rightarrow\infty
\end{align*}
as $l\rightarrow\infty$, since $||e_{(m_{l})}||^{2}\frac{1}{2}\sum_{i=1}^{n}(\theta_{2}^{\textrm{T}} x_{i})z_{i,o}^{2}\rightarrow\infty$
as $l\rightarrow\infty$. Therefore $-\mathscr{L}(e,\theta)$ grows
unboundedly as $||e||\rightarrow\infty$, and $\mathcal{B}_{c}(\theta)$
is bounded. Since $e\mapsto-\mathscr{L}(e,\theta)$ is continuous
over $\mathbb{R}^{n}$ for $\theta\in\Theta_{n}$, $\mathcal{B}_{c}(\theta)$
is also closed. It then follows from the Weierstrass theorem and $\Theta_{n}\subseteq\Theta_{0,n}$
that there exists $e(\theta)\in\arg\min_{e\in\mathbb{R}^{n}}\left\{-\mathscr{L}(e,\theta)\right\}=\arg\max_{e\in\mathbb{R}^{n}}\mathscr{L}(e,\theta)$,
for all $\theta\in\Theta_{n}$.

Step 2. The Hessian matrix of the map $e\mapsto\mathscr{L}(e,\theta)$
is $-\textrm{diag}[H_{x_{i}}'\{e(y_{i},x_{i},\theta);\theta\}]$,
and is thus negative definite for all $\theta\in\Theta_{n}$. Therefore,
$e\mapsto\mathscr{L}(e,\theta)$ is strictly concave with unique maximum
$e(\theta)$ for all $\theta\in\Theta_{n}$. %Define the function $L:\mathbb{R}^{K}\times\mathbb{R}\times\mathbb{R}^{2+J(K-1)}\rightarrow\mathbb{R}$
%as
%\begin{equation}
%L(x_{i},y_{i},\theta)=\sum_{j=2}^{J}(\theta_{j}^{\textrm{T}} x_{ij}^{c})\left[h_{j}\{e(y_{i},x_{i},\theta)\}e(y_{i},x_{i},\theta)-\widetilde{h}_{j}\{e(y_{i},x_{i},\theta)\}\right].%\label{eq:L(X,Y)}
%\end{equation}
With $L(x_{i},y_{i},\theta)$ defined in (\ref{eq:L(X,Y)}), and upon using first-order conditions (\ref{eq:FOCsGD}), direct substitution
yields $\mathscr{L}\left\{e(\theta),\theta\right\}=\sum_{i=1}^{n}L(x_{i},y_{i},\theta)$,
the maximum of the map $e\mapsto\mathscr{L}(e,\theta)$ for all $\theta\in\Theta_{n}$,
and the primal objective function.

\textit{Proof of part (ii).} By Lemma \ref{lem:convexity}, the
first-order conditions of (GP) implied by (\ref{eq:fock}) coincide
with the system
\begin{eqnarray}
\sum_{i=1}^{n}x_{ij}^{c}\widetilde{h}_{j}(e_{i})=0 \quad(j=1,\ldots,J), \quad y_{i}=\sum_{j=1}^{J}(\theta_{j}^{\textrm{T}} x_{ij}^{c})h_{j}(e_{i}) \quad(i=1,\ldots,n).\label{eq:GDR_MM}
\end{eqnarray}
 Moreover, the $n$ first-order conditions
(\ref{eq:FOCsGD}) and the constraints of (GD) together yield the
method-of-moments representation of (GD).

\textit{Proof of part (iii).} (a) By Condition \ref{ExistenceGDR},
there exists $\theta\in\Theta_{n}$ such that first-order conditions
(\ref{eq:GDR_MM}) are satisfied. By Lemma \ref{lem:convexity},
$Q_{n}(\theta)$ is strictly convex over $\Theta_{n}$. Therefore,
$\theta_{n}$ is the unique minimum of $Q_{n}(\theta)$ and uniquely
solves (\ref{eq:GDR_MM}).

By definition, a solution $e^{*}$ to (GD) with Lagrange multiplier
$\theta^{*}$ satisfies first-order conditions (\ref{eq:FOCsGD}).
Suppose $\theta^{*}\in\Theta_{n}$. By Step 2 in part (i), the map
$e\mapsto\mathscr{L}(e,\theta)$ admits a unique maximizer $e(\theta)$,
for all $\theta\in\Theta_{n}$: each pair $\left\{\theta,e(\theta)\right\}$ is
well-defined and satisfies first-order conditions (\ref{eq:FOCsGD}).
Since $\theta_{n}$ uniquely solves (\ref{eq:GDR_MM}) over $\Theta_{n}$,
the pair $\left\{\theta_{n},e(\theta_{n})\right\}$ is the unique pair satisfying
system (\ref{eq:GDR_MM}) in $\Theta_{n}\times\mathcal{E}_{n}$, where
$\mathcal{E}_{n}=\{e\in\mathbb{R}^{n}:y_{i}=H_{x_{i}}(e_{i};\theta)\quad(i=1,\ldots,n),\;\textrm{for some }\theta\in\Theta_{n}\}$
is the set of admissible optimal solutions to (GD). It follows that
the pair $(\theta^{*},e^{*})=\left\{\theta_{n},e(\theta_{n})\right\}$ is the unique
pair satisfying system (\ref{eq:GDR_MM}) in $\Theta_{n}\times\mathcal{E}_{n}$.
Therefore, the pair $(\theta_{n},e^{*})$ uniquely solves (GP) and
(GD) over $\Theta_{n}\times\mathcal{E}_{n}$.

Suppose $\theta^{*}\notin\Theta_{n}$. For $\theta\notin\Theta_{n}$,
a pair $\left\{\theta,e(\theta)\right\}$ (not necessarily unique) does not satisfy
the second-order conditions of (GD). Thus a solution to (GD) with
Lagrange multipliers $\theta^{*}\notin\Theta_{n}$ is not a global
maximum of (GD) over $\mathbb{R}^{n}$. Thus there is no
solution to (GD) such that both $\theta^{*}\notin\Theta_{n}$ and
the value of (GD) is equal to or exceeds the optimal value of (GD)
at  $e^{*}=e(\theta_{n})$. Therefore, the pair $\left\{\theta_{n},e^{*}\right\}$
is the unique optimal solution to (GP) and (GD) over $\Theta_{n}\times\mathbb{R}^{n}$.

\begin{sloppy}
(b) By direct substitution and using that $\sum_{i=1}^{n}x_{i}^{c}e_{i}^{*}=0$,
at a solution the value of (GD) is $\sum_{i=1}^{n}y_{i}e_{i}^{*}=\sum_{i=1}^{n}\sum_{j=2}^{J}(\theta_{j}^{*\textrm{T}} x_{ij}^{c})h_{j}(e_{i}^{*})e_{i}^{*}$.
Using that $\sum_{i=1}^{n}(\theta_{jn}^{\textrm{T}} x_{ij}^{c})\widetilde{h}_{j}\{e(y_{i},x_{i},\theta_{n})\}=0$ 
$(j=1,\ldots,J)$, at a solution the value of (GP) is $\sum_{i=1}^{n}\sum_{j=2}^{J}(\theta_{jn}^{\textrm{T}} x_{ij}^{c})h_{j}\{e(y_{i},x_{i},\theta_{n})\}e(y_{i},x_{i},\theta_{n})$.
Strong duality then follows from $\theta_{n}=\theta^{*}$ established
in (a). 
\par\end{sloppy}

\section{Proof of Theorem 4}
\begin{proof}
\textit{Proof of part (i).} Let $U$  satisfy
$U\sim U(0,1)$ and $E(X\mid U)=E(X)$. Then:
\[
E\{X\otimes m^{J}(U)\}=E\{E(X\mid U)\otimes m^{J}(U)\}=E\{E(X)\otimes m^{J}(U)\}=E(X)\otimes E\{m^{J}(U)\}=0
\]
for all $J$. The first equality holds by iterated expectations, the
second by mean independence, the third by linearity of the expectation,
and the last by uniformity of $U$ and definition of $m^{J}$.

In order to show the converse statement, suppose that $E(X\mid U)=E(X)$
does not hold. Following steps similar to the proof of Lemma 2.1 in
\citet{DIN:2003}, and letting $\varphi(U)=E(X\mid U)-E(X)$, for $\varPsi_{J}$
such that $E[||\varphi(U)-\varPsi_{J}m^{J}(U)||^{2}]\rightarrow0$,
\[
E\left[m^{J}(U)^{\textrm{T}}\varPsi_{J}^{\textrm{T}}\{X-E(X)\}\right]=E\left[m^{J}(U)^{\textrm{T}}\varPsi_{J}^{\textrm{T}}\{E(X\mid U)-E(X)\}\right]\rightarrow E[||\varphi(U)||^{2}]>0,
\]
as $J\rightarrow\infty$, which implies $E[X\otimes m^{J}(U)]\neq0$
for all $J$ large enough, since
\[
E\left[m^{J}(U)^{\textrm{T}}\varPsi_{J}^{\textrm{T}}\{X-E(X)\}\right]=E[\{X-E(X)\}\otimes m^{J}(U)]\textrm{vec}(\varPsi_{J})=E\{X\otimes m^{J}(U)\}\textrm{vec}(\varPsi_{J}).
\]
Now suppose that $U\sim U(0,1)$ does not hold. Because $X$ includes
an intercept, any random variable $\widetilde{U}$ such that $E\{X\otimes m^{J}(\widetilde{U})\}=0$
for all $J$ must also satisfy $E\{m^{J}(\widetilde{U})\}=0$ for all
$J$, and therefore $\widetilde{U}\sim U(0,1)$. It follows that $E\{X\otimes m^{J}(U)\}\neq0$
in the large $J$ limit.

Therefore, $E\{X\otimes m^{J}(U)\}=0$ for all $J$ if and only if
$E(X\mid U)=E(X)$ and $U\sim U(0,1)$, and the result follows.

\textit{Proof of part (ii).} Let $e$ be a random variable with mean
$0$ and variance $1$ satisfying $E(\widetilde{X}\mid e)=0$. Then $E\{\widetilde{X}^{c}\otimes\widetilde{h}^{J}(e)\}=E\{E(\widetilde{X}^{c}\mid e)\otimes\widetilde{h}^{J}(e)\}=0$,
for all $J$, by iterated expectations and mean independence.

In order to show the converse statement, suppose that $E(\widetilde{X}^{c}\mid e)\neq0$.
Letting $\varphi(e)=E(\widetilde{X}^{c}\mid e)$ and $\varPsi_{J}$ such that
$E\{||\varphi(e)-\varPsi_{J}\widetilde{h}^{J}(e)||^{2}\}\rightarrow0$,
and following steps similar to the proof of Lemma 2.1 in \citet{DIN:2003},
\[
E\{\widetilde{X}^{c}\otimes\widetilde{h}^{J}(e)\}\textrm{vec}(\varPsi_{J})=E\{\widetilde{h}^{J}(e)^{\textrm{T}}\varPsi_{J}^{\textrm{T}}\widetilde{X}^{c}\}=E\{\widetilde{h}^{J}(e)^{\textrm{T}}\varPsi_{J}^{T}E(\widetilde{X}^{c}\mid e)\}\rightarrow E\{||\varphi(e)||^{2}\}>0,
\]
as $J\rightarrow\infty$, which implies $E\{\widetilde{X}^{c}\otimes\widetilde{h}^{J}(e)\}\neq0$
as $J\rightarrow\infty$.

Therefore, a random variable $e$ with mean $0$ and variance $1$
satisfies $E\{\widetilde{X}^{c}\otimes\widetilde{h}^{J}(e)\}=0$ for all
$J$ large enough if and only if $E(\widetilde{X}^{c}\mid e)=0$, and the
result follows.
\end{proof}

\section{Asymptotic Theory}

In this Section, $C$ denotes a generic constant whose value may vary
from place to place.

\subsection{Proof of Theorem 5}

Letting $e=e(Y,X,\theta)$ for $\theta\in\Theta$, by definition (\ref{eq:L(X,Y)}),
$L(X,Y,\theta)$ can be decomposed as 
\begin{equation}
L(X,Y,\theta)=\frac{1}{2}(\theta_{2}^{\textrm{T}} X^{c})(e^{2}+1)+\sum_{j=3}^{J}(\theta_{j}^{\textrm{T}} \widetilde{X}^{c})\{h_{j}(e)e-\widetilde{h}_{j}(e)\}\equiv L_{1}(X,Y,\theta)+L_{2}(X,Y,\theta).\label{eq:L1L2}
\end{equation}
Define $Q_{0}(\theta)=E\left\{L(X,Y,\theta)\right\}$, the population objective
of the generalized primal problem.

Both existence and consistency of $\hat{\theta}$ result from strict
convexity of $Q_{0}(\theta)$, and pointwise convergence of $Q_{n}(\theta)$
to $Q_{0}(\theta)$, since strict convexity and pointwise convergence
together imply uniform convergence, as in, for instance, Theorem 2.7
in \citet{Newey:McFadden:1994}. The asymptotic distribution of $\hat{\theta}$
follows from the method-of-moments characterization of the estimates
given in part (ii) of Theorem 3, and Theorem
3.4 in \citet{Newey:McFadden:1994}.

\textit{Proof of parts (i) and (ii)}.We verify the
conditions of Theorem 2.7 in \citet{Newey:McFadden:1994}. We first
show that $\theta_{0}$ is the unique minimizer of $Q_{0}(\theta)$
in $\Theta$, using the next result.
%\begin{lemmaS4*}
\begin{lemma}
\label{lem:QDiff}Suppose that Conditions \ref{Data}, \ref{Design},
\ref{FR}, \ref{LimFR} and \ref{Moments}(i) hold. Then, $Q_{0}(\theta)$
is continuously differentiable, $E\{|L(X,Y,\theta)|\}<\infty$ and $\nabla_{\theta}E\{L(X,Y,\theta)\}=E\{\nabla_{\theta}L(X,Y,\theta)\}$
for $\theta\in\Theta$.
\label{lem:QDiff}
\end{lemma}
%\end{lemmaS4}
\begin{sloppy}
\begin{proof}
We first show that $E\{|L(X,Y,\theta)|\}<\infty$ for all $\theta\in\Theta$.
$|L_{1}|$ in (\ref{eq:L1L2}) satisfies
\begin{equation}
\left|\frac{1}{2}(\theta_{2}^{\textrm{T}} X^{c})(e^{2}+1)\right|\leq\frac{1}{2}||\theta_{2}||\,||X^{c}||\,(e^{2}+1),\label{eq:Bound1}
\end{equation}
which has finite expectation if $E(||X^{c}||e^{2})$ and $E(||X^{c}||)$
are bounded. Since $\{h_{j}\}_{j=3,\ldots,J}$ and $\{\widetilde{h}_{j}\}_{j=3,\ldots,J}$
are bounded, $|L_{2}|$ in (\ref{eq:L1L2}) satisfies
\begin{equation}
\left|\sum_{j=3}^{J}(\theta_{j}^{\textrm{T}} \widetilde{X}^{c})\{h_{j}(e)e-\widetilde{h}_{j}(e)\}\right|\leq C\sum_{j=3}^{J}||\theta_{j}||(||X^{c}||\,|e|+||X^{c}||),\label{eq:Bound2}
\end{equation}
which has finite expectation if $E(||X^{c}||\,|e|)$ and $E(||X^{c}||)$
are bounded. It follows that $|L(X,Y,\theta)|$ has finite expectation
if $E(||X^{c}||e^{2})<\infty$.

The identity $Y=\theta_{1}^{\textrm{T}} X^{c}+(\theta_{2}^{\textrm{T}} X^{c})e+\sum_{j=3}^{J}(\theta_{j}^{\textrm{T}} \widetilde{X}^{c})h_{j}(e)$
holds with probability one for $\theta\in\Theta$, and $\{h_{j}\}_{j=3,\ldots,J}$
bounded thus implies
\begin{align}
|e^{2}| & =|(\theta_{2}^{\textrm{T}} X^{c})^{-2}\{Y-\theta_{1}^{\textrm{T}} X^{c}-\sum_{j=3}^{J}(\theta_{j}^{\textrm{T}} \widetilde{X}^{c})h_{j}(e)\}^{2}|\nonumber \\
 & \leq C\{\inf_{x\in\mathcal{X}}(\theta_{2}^{\textrm{T}} x^{c})\}^{-2}\{2|Y|^{2}+2(||\theta_{1}||^{2}\,||X^{c}||^{2}+\sum_{j=3}^{J}||\theta_{j}||^{2}\,||X^{c}||^{2})\}.\label{eq:Boundesq}
\end{align}
Therefore,
\[
E(||X^{c}||\,|e^{2}|)\leq C\{\inf_{x\in\mathcal{X}}(\theta_{2}^{\textrm{T}} x^{c})\}^{-2}E(||X^{c}||\,|Y|^{2}+||\theta_{1}||^{2}\,||X^{c}||^{3}+\sum_{j=3}^{J}||\theta_{j}||^{2}\,||X^{c}||^{3})<\infty.
\]
Bounds (\ref{eq:Bound1}) and (\ref{eq:Bound2}) now imply $E\{|L_{1}(X,Y,\theta)|\}<\infty$
and $E\{|L_{2}(X,Y,\theta)|\}<\infty$, for all $\theta\in\Theta$ since
$\Theta$ is bounded. Hence $E\{|L(X,Y,\theta)|\}<\infty$ for all
$\theta\in\Theta$. 

Bound (\ref{eq:Boundesq}) implies that $E\{\sup_{\theta\in\Theta}||\nabla_{\theta}L(X,Y,\theta)||\}<\infty$.
By Lemma \ref{lem:convexity}, $\nabla_{\theta_{1}}L(X,Y,\theta)=-X^{c}e$
and $\nabla_{\theta_{2}}L(X,Y,\theta)=-X^{c}(e^{2}-1)/2$, and $\nabla_{\theta_{j}}L(X,Y,\theta)=-\widetilde{X}^{c}\widetilde{h}_{j}(e)$.
Bound (\ref{eq:Boundesq}) together with $\{\widetilde{h}_{j}\}_{j=3,\ldots,J}$
bounded, boundedness of $\Theta$ and Holder's inequality thus imply
that $E\left\{\sup_{\theta\in\Theta}||\nabla_{\theta}L(X,Y,\theta)||\right\}<\infty$
under Condition \ref{Moments}(i). Lemma 3.6 in \citet{Newey:McFadden:1994}
then implies that $Q_{0}(\theta)$ is continuously differentiable
and that the order of differentiation and integration can be interchanged
for $\theta\in\Theta$.
\end{proof}
By Lemma \ref{lem:QDiff}, $Q_{0}(\theta)$ is continuously differentiable
and the order of differentiation and integration can be interchanged,
for $\theta\in\Theta$. Moreover, $\nabla_{\theta}Q_{0}(\theta)$
is differentiable for $\theta\in\Theta$. Letting $P=(1,e)^{\textrm{T}}$
and $Q=\{h_{3}(e),\ldots,h_{J}(e)\}^{\textrm{T}}$, from the proof of
Lemma \ref{lem:convexity}, 
\[
\nabla_{\theta\theta}L(X,Y,\theta)=\left[\begin{array}{cc}
\frac{X^{c}X^{c\textrm{T}}}{H_{X}'(e;\theta)}\otimes PP^{\textrm{T}} & \frac{X^{c}\widetilde{X}^{c\textrm{T}}}{H_{X}'(e;\theta)}\otimes PQ^{\textrm{T}}\\
\frac{\widetilde{X}^{c}X^{c\textrm{T}}}{H_{X}'(e;\theta)}\otimes QP^{\textrm{T}} & \frac{\widetilde{X}^{c}\widetilde{X}^{c\textrm{T}}}{H_{X}'(e;\theta)}\otimes QQ^{\textrm{T}}
\end{array}\right].
\]
Applying steps similar to those leading to the bound (\ref{eq:Boundesq})
in the proof of Lemma \ref{lem:QDiff} and using that $\inf_{e\in\mathbb{R}}H_{X}'(e;\theta)>0$
for all $\theta\in\Theta$ shows that $||\{X^{c}X^{c\textrm{T}}/H_{X}'(e;\theta)\}e^{2}||$
has finite expectation for all $\theta\in\Theta$ under Condition
\ref{Moments}(i). Therefore, boundedness of $\{h_{j}\}_{j=3,\ldots,J}$
and $\Theta$, and Holder's inequality imply that $E\left\{\sup_{\theta\in\Theta}||\nabla_{\theta\theta}L(X,Y,\theta)||\right\}<\infty$.
It then follows from Lemma 3.6 in \citet{Newey:McFadden:1994} that
$\nabla_{\theta}Q_{0}(\theta)$ is continuously differentiable, and
that the Hessian matrix of $Q_{0}(\theta)$ is $H(\theta)=E\left\{\nabla_{\theta\theta}L(X,Y,\theta)\right\}$,
which is a finite positive definite matrix under the assumed conditions,
by application of Lemma \ref{lem:convexity}. Therefore, $Q_{0}(\theta)$
is strictly convex and $\theta_{0}$ is the unique minimizer of $Q_{0}(\theta)$
in $\Theta$, and Condition (i) of Newey and McFadden's Theorem 2.7
is verified.

By Condition \ref{DGP}, $\theta_{0}$ is in the interior of $\Theta$,
which is convex, and $Q_{n}(\theta)$ is convex with probability 1
by Lemma \ref{lem:convexity}, and their Condition (ii) is verified.
Finally, since the sample is independently and identically distributed by assumption, pointwise convergence of $Q_{n}(\theta)$
to $Q_{0}(\theta)$ follows from boundedness of $Q_{0}(\theta)$, established
in the proof of Lemma \ref{lem:QDiff}, and application of Khinchine's
law of large numbers. All conditions of Newey and McFadden's
Theorem 2.7 are therefore satisfied, and there exists $\hat{\theta}\in\Theta$
with probability approaching one and $\hat{\theta}$ converges in
probability to $\theta_{0}$.

\textit{Proof of part (iii).} Define $m_{j}(Y,X,\theta)=X^{c}\widetilde{h}_{j}\{e(Y,X,\theta)\}$
for $j=1,2$, and $m_{j}(Y,X,\theta)=\widetilde{X}^{c}\widetilde{h}_{j}\{e(Y,X,\theta)\}$
for $j=3,\ldots,J$, and let $m(Y,X,\theta)=\{m_{1}(Y,X,\theta),\ldots,m_{J}(Y,X,\theta)\}^{\textrm{T}}$,
$G=E\{\nabla_{\theta}m(Y,X,\theta)\}|_{\theta=\theta_{0}}$ and $S=E\{m(Y,X,\theta_{0})m(Y,X,\theta_{0})^{\textrm{T}}\}$.
By part (ii) of Theorem 3, the Lagrange
multiplier vector $\hat{\theta}$ solves the $2+J(K-1)$ equations
system 
\begin{equation}
\frac{1}{n}\sum_{i=1}^{n}m(y_{i},x_{i},\theta)=0.\label{eq:system}
\end{equation}
System (\ref{eq:system}) can be equivalently viewed as minimizing
\begin{eqnarray*}
\mathcal{Q}_{n}^{MM}(\theta) & = & \left\{ \frac{1}{n}\sum_{i=1}^{n}m(y_{i},x_{i},\theta)\right\} ^{\textrm{T}}\left\{ \frac{1}{n}\sum_{i=1}^{n}m(y_{i},x_{i},\theta).\right\} 
\end{eqnarray*}
Asymptotic normality of the method-of-moments estimator then follows
after verifying conditions of Theorem 3.4 in \citet{Newey:McFadden:1994}.

From the proof of Lemma \ref{lem:convexity}, the derivative of
$e$ with respect to $\theta_{j}$ is $\nabla_{\theta_{j}}e=-X^{c}h_{j}(e)\{H_{X}'(e;\theta)\}^{-1}$,
for $j=1,2$, and $\nabla_{\theta_{j}}e=-\widetilde{X}^{c}h_{j}(e)\{H_{X}'(e;\theta)\}^{-1}$,
for $j=3,\ldots,J$, which is continuous for all $\theta\in\Theta$
with probability one by definition of $e$ and $\Theta$. Thus the
mapping $\theta\mapsto m(Y,X,\theta)$ is continuously differentiable
in $\theta\in\Theta$ with probability one, and Newey and McFadden's
Condition (ii) is satisfied. By definition $\varepsilon=e(Y,X,\theta_{0})$
is independent of $X$ which implies that $E\{m(Y,X,\theta_{0})\}=0$,
and the first part of their Condition (iii) is satisfied. In addition,
steps similar to the proof of Lemma \ref{lem:QDiff} show that $E\{||m(Y,X,\theta_{0})||^{2}\}$
and $E\{\sup_{\theta\in\varTheta}||\nabla_{\theta}m(Y,X,\theta)||\}$
are finite under Conditions \ref{Moments}, and their Conditions (iii)--(iv)
are thus verified. Finally, their full rank condition on $G=E\{\nabla_{\theta}m(Y,X,\theta)\}|_{\theta=\theta_{0}}$
is satisfied under our conditions since $G=H(\theta_{0})$ is then
positive definite. Therefore, $n^{1/2}(\hat{\theta}-\theta_{0})$
converges in distribution to $N(0,\Sigma)$ with
\begin{equation}
\Sigma=G^{-1}S(G^{-1})^{\textrm{T}}.
\end{equation}

\par\end{sloppy}

\subsection{Proof of Theorem 6}

\begin{sloppy}
\textit{Proof of part (i). }Denote the cumulative distribution function
of $\hat{e}=e(Y,X,\hat{\theta})$, by $\hat{F}(e)=E\{1(\hat{e}\leq e)\}$
for all $e\in\mathbb{R}$, and consider the decomposition
\begin{equation}
F_{n}(e)-F(e)=\{F_{n}(e)-\hat{F}(e)\}+\{\hat{F}(e)-F(e)\}\quad\quad(e\in\mathbb{R}).\label{eq:decompo1-1}
\end{equation}
For the first term, convergence in probability of $\sup_{e}|\hat{F}(e)-F(e)|$
to 0 is implied by Glivenko-Cantelli (e.g., Theorem 19.1 in \citealp{vdV:1998}).
For the second term, upon using that the events $\{e(Y,X,\theta)\leq e\}$
and $\{Y\leq H_{X}(e;\theta)\}$ are equivalent conditional on $X$
for $\theta\in\Theta$, and in particular for $\hat{\theta},\theta_{0}\in\Theta$,
applying iterated expectations, a change of variable and a mean-value
expansion, yields
\begin{eqnarray*}
\hat{F}(e)-F(e) & = & E[E\{1(\hat{e}\leq e)\mid X\}-E\{1(\varepsilon\leq e)\mid X\}]\\
 & = & E[F_{Y\mid X}\{H_{X}(e;\hat{\theta})\mid X\}-F_{Y\mid X}\{H_{X}(e;\theta_{0})\}\mid X]\\
 & = & (\hat{\theta}-\theta_{0})^{\textrm{T}}E[f_{Y\mid X}\{H_{X}(e;\bar{\theta})\mid X\}m\{H_{X}(e;\theta),X,\theta\}],
\end{eqnarray*}
where $\bar{\theta}$ is on the line connecting $\hat{\theta}$ and
$\theta_{0}$. Since $\sup_{e}e^{2}f_{Y\mid X}\{H_{X}(e;\theta)\mid X\}$
and $\{\widetilde{h}_{j}\}_{j=3,\ldots,J}$ are uniformly bounded, it
follows that 
\[
\sup_{e\in\mathbb{R}}|\hat{F}(e)-F_{\varepsilon}(e)|\leq C||\hat{\theta}-\theta_{0}||E(||X^{c}||).
\]
Consistency of $\hat{\theta}$ and $E(||X^{c}||)$ finite then imply
convergence in probability of $\sup_{e}|\hat{F}(e)-F(e)|$ to 0. The
result follows from combining the two uniform convergence results.

\textit{Proof of part (ii).} For $D=(Y,X)$, let $\mathbb{E}_{n}f=\mathbb{E}_{n}f(d_{i})=n^{-1}\sum_{i=1}^{n}f(d_{i})$
and $\mathbb{G}_{n}f=\mathbb{G}_{n}\{f(d_{i})\}=n^{-1/2}\sum_{i=1}^{n}[f(d_{i})-E\{f(d_{i})\}]$,
and define the class of functions $\mathcal{F}=\left\{ 1\{e(Y,X,\theta)\leq e\},e\in\mathbb{R},\theta\in\Theta\right\} $.
Following \citet{vdVW:2007}, the empirical dual regression process
$\mathbb{U}_{n}(e)=n^{1/2}(\mathbb{E}_{n}f_{e,\hat{\theta}}-Ef_{e,\theta_{0}})$
admits the following decomposition: 
\begin{equation}
n^{1/2}(\mathbb{E}_{n}f_{e,\hat{\theta}}-Ef_{e,\theta_{0}})=\mathbb{G}_{n}(f_{e,\hat{\theta}}-f_{e,\theta_{0}})+\mathbb{G}_{n}f_{e,\theta_{0}}+\sqrt{n}E(f_{e,\hat{\theta}}-f_{e,\theta_{0}}).\label{eq:Decompo}
\end{equation}
The proof thus proceeds by (i) establishing that the first term on
the right in (\ref{eq:Decompo}) converges in probability to zero,
(ii) using the fact that the second term converges in distribution
to a mean zero Gaussian process, and (iii) expanding the last term
uniformly in $e\in\mathbb{R}$.
\par\end{sloppy}

Step 1. Stochastic equicontinuity. By Theorem 2.1 in \citet{vdVW:2007},
since $\Pr(\hat{\theta}\in\Theta)\rightarrow1$ by part (i) of Theorem
5, $\sup_{e\in\mathbb{R}}|\mathbb{G}_{n}(f_{e,\hat{\theta}}-f_{e,\theta_{0}})|$
converges in probability to $0$ holds if the class of functions $\mathcal{F}$
is Donsker and if the pseudometric $\rho\{(e',\theta'),(e'',\theta'')\}^{2}\equiv E[\{f_{e',\theta'}(D)-f_{e'',\theta''}(D)\}^{2}]$
satisfies $\delta_{n}\equiv\sup_{e\in\mathbb{R}}\rho\{(e,\hat{\theta}),(e,\theta_{0})\}^{2}$
converges in probability to 0.

We first show that the class of functions $\mathcal{F}$ is Donsker.
Define the parametric class of functions $\widetilde{\mathcal{F}}=\{e(Y,X,\theta),\theta\in\Theta\}$.
For all $\theta',\theta''\in\Theta$, a mean-value expansion and Cauchy-Schwarz
inequality yield 
\[
|e(Y,X,\theta')-e(Y,X,\theta'')|\leq||\nabla_{\theta}e(Y,X,\theta)|_{\theta=\bar{\theta}}||\,||\theta'-\theta''||,
\]
where $\bar{\theta}$ is on the line joining $\theta'$ and $\theta''$.
Steps similar to those in the proof of Theorem 5
show that $E\{||\nabla_{\theta}e(Y,X,\theta)|_{\theta=\bar{\theta}}||^{2}\}$
is bounded under Condition \ref{Moments}, so that $\widetilde{\mathcal{F}}$
is Donsker by Example 19.7 in \citet{vdV:1998}. Therefore, $\mathcal{F}$
is Donsker, by monotonicity of the indicator function, with unit envelope.

We now show that $\delta_{n}$ converges in probability
to $0$. Since the events $\{e(Y,X,\theta)\leq e\}$ and $\{Y\leq H_{X}(e;\theta)\}$
are equivalent conditional on $X$ for $\theta\in\Theta$, the law
of iterated expectations, a mean-value expansion and Cauchy-Schwarz
inequality yield: 
\begin{align*}
\sup_{e\in\mathbb{R}}\rho\{(e,\hat{\theta}),(e,\theta_{0})\}^{2} & =\sup_{e\in\mathbb{R}}E[|1\{e(Y,X,\hat{\theta})\leq e\}-1\{e(Y,X,\theta_{0})\leq e\}|]\\
 & =\sup_{e\in\mathbb{R}}E(|(\hat{\theta}-\theta_{0})^{\textrm{T}}[f_{Y\mid X}\{H_{X}(e;\bar{\theta})\mid X\}m\{H_{X}(e;\theta),X,\theta\}]|)\\
 & \leq C||\hat{\theta}-\theta_{0}||E(||X^{c}||),
\end{align*}
where $\bar{\theta}$ is on the line joining $\hat{\theta}$ and $\theta_{0}$.
Convergence in probability of $\delta_{n}$ to zero now follows from
$E(||X^{c}||)$ finite and consistency of $\hat{\theta}$.

Step 2. Expansion. Letting $g(e)=E[f_{Y\mid X}\{H_{X}(e;\theta_{0})\mid x_{i}\}m\{H_{X}(e;\theta_{0}),X,\theta_{0}\}]$,
$e\in\mathbb{R}$, we show that the following expansion is valid uniformly
in $e\in\mathbb{R}$: 
\begin{align}
E\{f_{e,\hat{\theta}}(D)-f_{e,\theta_{0}}(D)\} & =(\hat{\theta}-\theta_{0})^{\textrm{T}}\{g(e)+o_{P}(1)\}.\label{eq:Expansion1-1}
\end{align}
Steps similar to above yield: 
\[
E\{f_{e,\hat{\theta}}(D)-f_{e,\theta_{0}}(D)\}=(\hat{\theta}-\theta_{0})^{\textrm{T}}E[\nabla_{\theta}F_{Y\mid X}\{H_{x_{i}}(e;\theta)\mid x_{i}\}|_{\theta=\bar{\theta}}],
\]
where $\bar{\theta}$ is on the line joining $\hat{\theta}$ and $\theta_{0}$.
We obtain 
\[
E[f_{Y\mid X}\{H_{X}(e;\bar{\theta})\mid x_{i}\}m\{H_{X}(e;\theta),X,\theta\}]=E[f_{Y\mid X}\{H_{X}(e;\theta_{0})\mid X\}m\{H_{X}(e;\theta),X,\theta\}]+o_{p}(1),
\]
uniformly in $e\in\mathbb{R}$, by uniform continuity of the mapping
$y\mapsto f_{Y\mid X}(y\mid x)$, uniformly in $x$ over $\mathcal{X}$, consistency
of $\hat{\theta}$, and since $\sup_{e\in\mathbb{R}}e^{2}f_{Y\mid X}\{H_{X}(e;\theta_{0})\mid X\}$,
$\max_{j=3,\ldots,J}\{\widetilde{h}_{j}\}$ are bounded and $E(||X^{c}||)$
is finite. Hence (\ref{eq:Expansion1-1}) holds by definition of $g(e)$,
uniformly in $e\in\mathbb{R}$.

Finally, the method-of-moments representation of dual
regression implies that the dual regression estimator $\hat{\theta}$
is asymptotically linear with influence function 
\begin{equation}
\psi(Y,X,\theta_{0})=-G^{-1}m(Y,X,\theta_{0}).\label{eq:Influence function}
\end{equation}
Thus (\ref{eq:Decompo})--(\ref{eq:Influence function}) together imply
that uniformly in $e\in\mathbb{R}$ 
\begin{align*}
\mathbb{U}_{n}(e) & =\mathbb{G}_{n}(f_{e,\hat{\theta}}-f_{e,\theta_{0}})+\mathbb{G}_{n}f_{e,\theta_{0}}+n^{1/2}(\hat{\theta}-\theta_{0})^{\textrm{T}}\{g(e)+o_{P}(1)\}\\
 & =o_{P}(1)+\mathbb{G}_{n}f_{e,\theta_{0}}+g(e)^{\textrm{T}}n^{-1/2}\sum_{i=1}^{n}\psi(y_{i},x_{i},\theta_{0})+o_{P}(1)\\
 & =n^{-1/2}\sum_{i=1}^{n}\varphi_{e}(y_{i},x_{i},\theta_{0})+o_{P}(1),
\end{align*}
where 
\[
\varphi_{e}(y_{i},x_{i},\theta_{0})=1\{e(y_{i},x_{i},\theta_{0})\leq e\}-F(e)-g(e)^{\textrm{T}}G^{-1}m(y_{i},x_{i},\theta_{0}).
\]
Therefore, the empirical dual regression process $\mathbb{U}_{n}$
weakly converges to the zero-mean Gaussian process $\mathbb{U}$,
where $\mathbb{U}$ has covariance function 
\begin{equation}
E\{\varphi_{e}(Y,X,\theta_{0})\varphi_{e'}(Y,X,\theta_{0})\}.
\end{equation}

%\newpage{}

\section{Numerical Illustrations}

\subsection{Implementation of generalized dual regression}

Define the criterion
\[
\textrm{SIC}(J,n,\theta)=2\sum_{i=1}^{n}\sum_{j=2}^{J}(\theta_{j}^{\textrm{T}} x_{ij}^{c})\{h_{j}(e_{i})e_{i}-\widetilde{h}_{j}(e_{i})\}+\{2+J(K-1)\}\log n,
\]
where $e_{i}$ solves $y_{i}=\sum_{j=1}^{J}(\theta_{j}^{\textrm{T}} x_{ij}^{c})h_{j}(e_{i})$ 
 $(i=1,\ldots,n)$, denoted $e(y_{i},x_{i},\theta)$. For $\theta^{*}$
such that $H_{x_{i}}'\{e(y_{i},x_{i},\theta^{*});\theta^{*}\}>0$
holds for each $i=1,\ldots,n$, using the strong duality result of
Theorem 3, the value of the criterion
can be computed as $y^{\textrm{T}}e^{*}+\{2+J(K-1)\}\log n$, with
$e^{*}$ the solution of the corresponding dual problem. We then select
an even value of $J$ according to the following alogorithm:

Step 1. For each $J$ in the grid $\{2,4,6,8\}$:

Step 1.1. Run program (GD) with basis functions specified as 
\[
h_{j}(e)=\begin{cases}
\cos\{2\pi(j-2)e\} & j\,\;\textrm{odd}\\
\sin\{2\pi(j-2)e\} & j\,\;\textrm{even}
\end{cases}\quad\quad(e\in\mathbb{R}),
\]
for $j=3,\ldots,J$, and $J$ even. Denote the solution by $e(J)$,
with corresponding multipliers $\theta(J)$.

Step 1.2. Compute $\textrm{SIC}\{J,n,\theta(J)\}=y^{\textrm{T}}e(J)+\{2+J(K-1)\}\log n$.

Step 2. Select the value of $J$ that minimizes $\textrm{SIC}\{J,n,\theta(J)\}$,
denoted $J^{*}$.

\begin{figure}
\subfloat[]{\includegraphics[width=7cm,height=8cm]{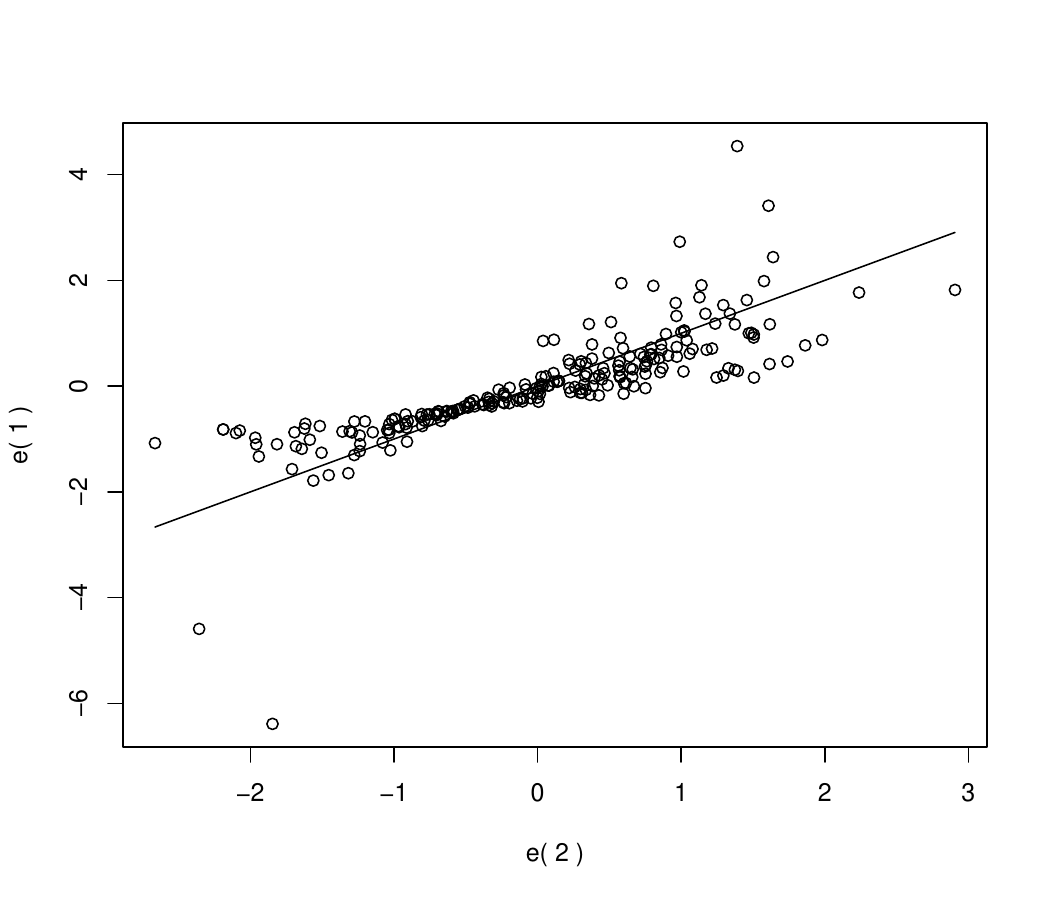}

}\hfill{} \subfloat[]{\includegraphics[width=7cm,height=8cm]{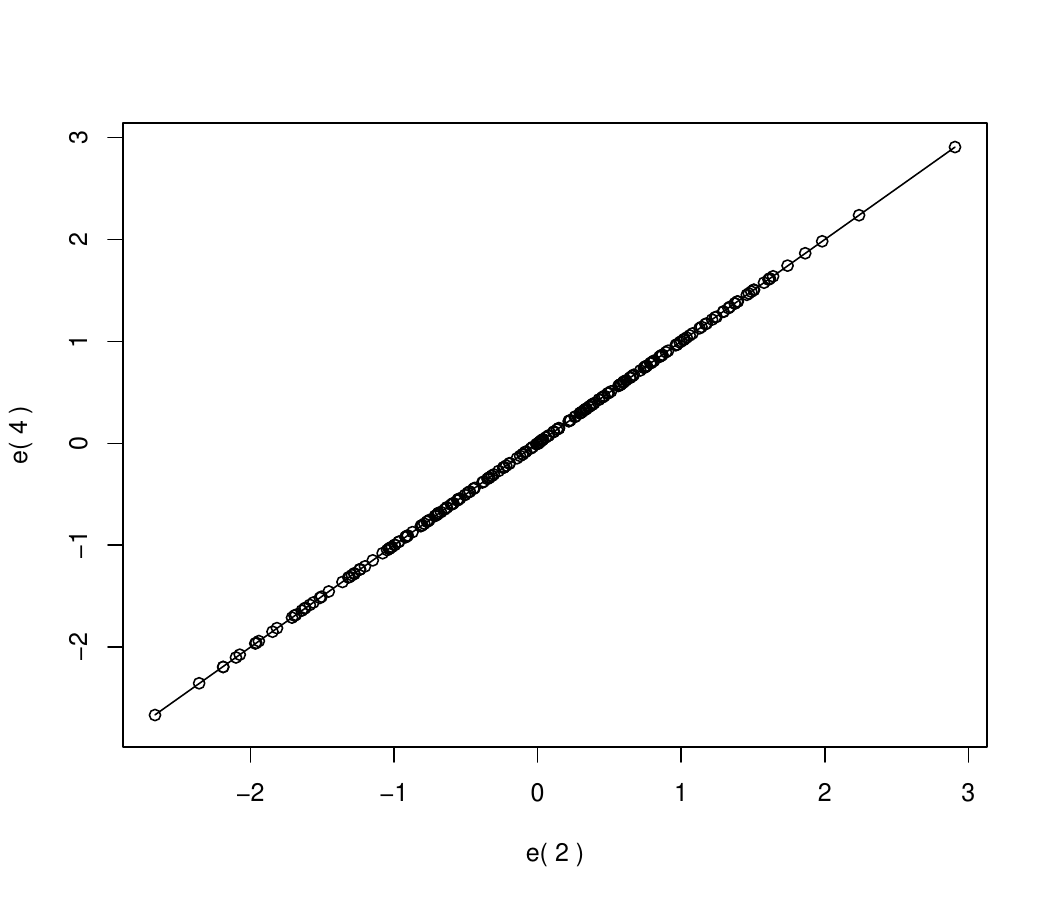}

}\hfill{}

\subfloat[]{\includegraphics[width=7cm,height=8cm]{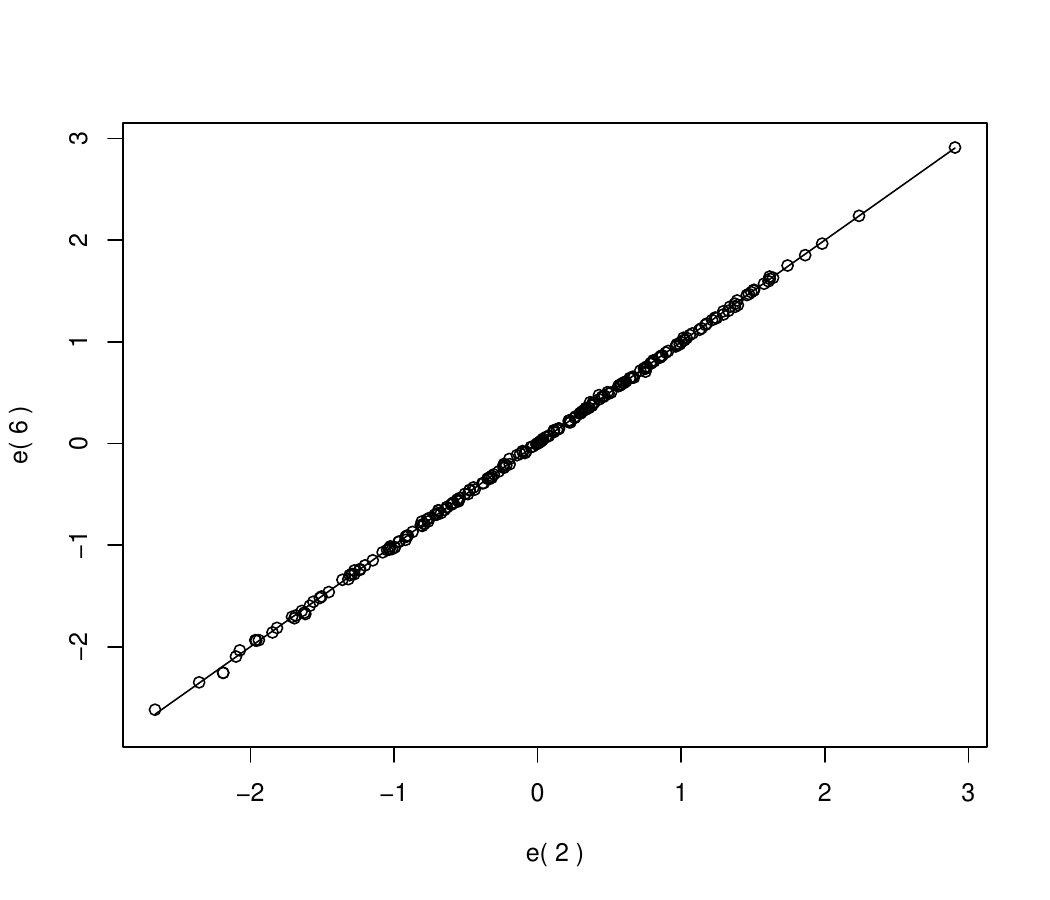}

}\hfill{} \subfloat[]{\includegraphics[width=7cm,height=8cm]{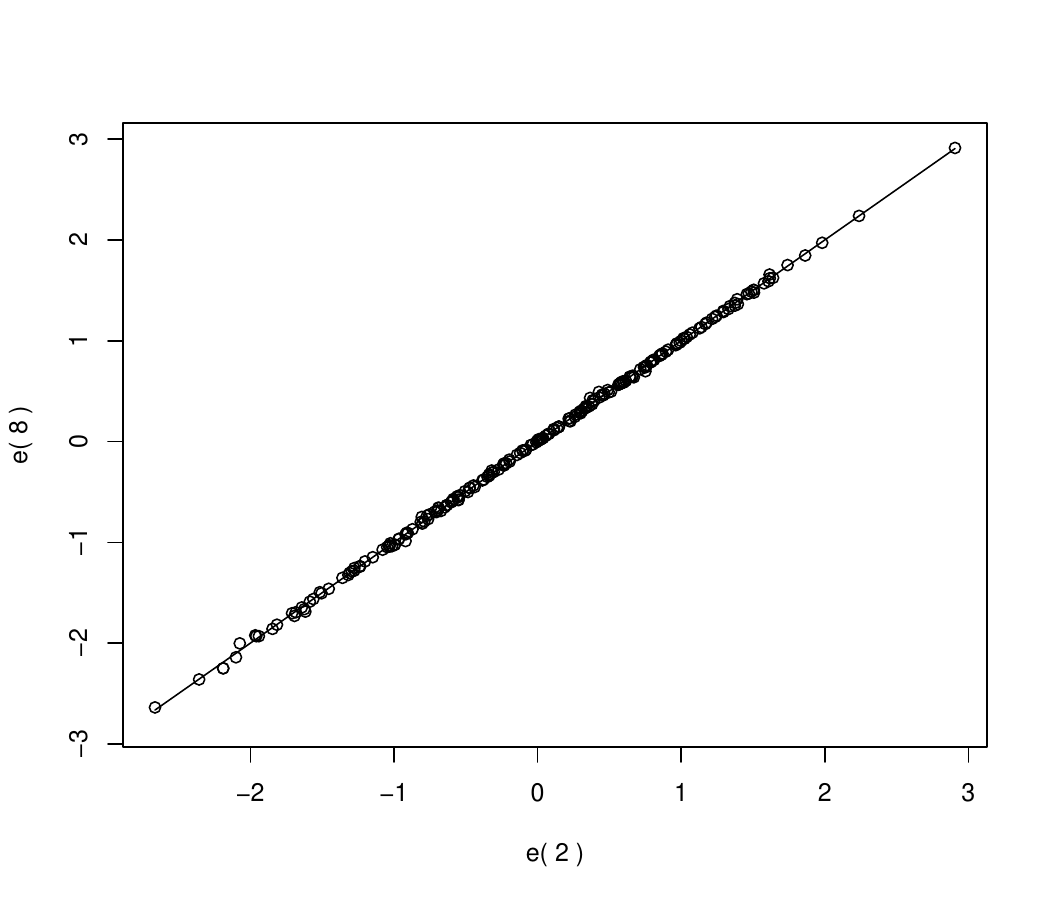}

}\hfill{}

\caption{$e(1)$, $e(4)$, $e(6)$ and $e(8)$ plotted against dual regression
solutions $e(2)$.\label{fig:edevs}}
\end{figure}

Results in the empirical application are robust to using a larger
grid for $J$. The grid specified above is also used in all simulations.
Although the proposed algorithm provides a convenient semi-automated
method for the specification of a generalized dual regression representation,
it is also instructive to examine the solutions obtained for $J$
greater than two. Fig. \ref{fig:edevs} plots the solutions $e(4)$,
$e(6)$ and $e(8)$ obtained in Step 1.1. against the selected solution
$e(2)$ in Step 2. We also plot the solution obtained for a location
model, denoted $e(1)$ and obtained as $F_{n}\{(y_{i}-\hat{\gamma}_{1}-\hat{\lambda}_{1}x_{i}^{c})/\hat{\gamma}_{2}\}$.
Visual inspection then confirms that although the dual regression
solution $e(2)$ differs significantly from the location solution
$e(1)$, our results are robust to the addition of extra terms in
the representation.

\subsection{Design and implementation of the numerical simulations\label{sec:Design-and-implementation}}

\begin{table}
\def~{\hphantom{0}}
\caption{Distribution of selected models across simulations in percentages.}{%
\begin{tabular}{llcccc}
& $n=100$ & $n=235$ & $n=500$ & $n=1000$\\[5pt]
$J^{*}=2$ &  $74\cdot97$ & $84\cdot14$ & $87\cdot70$ & $90\cdot32$\\
$J^{*}=4$ &  $24\cdot00$ & $15\cdot80$ & $12\cdot30$ & $9\cdot68$\\
$J^{*}=6$ &  $0\cdot72$ & $0\cdot06$ & $0\cdot00$ & $0\cdot00$\\
$J^{*}=8$ &  $0\cdot30$ & $0\cdot00$ & $0\cdot00$ & $0\cdot00$
\end{tabular}}
\label{tab:BICtab}
\end{table}

We generate $4999$ datasets of size $n=100,235,500,1000$ according
to the model $y_{i}=\alpha_{1}+\beta_{1}\widetilde{x}_{i}+(\alpha_{2}+\beta_{2}\widetilde{x}_{i})\varepsilon_{i}$
with $\varepsilon_{i}\sim N(0,1)$ and $\widetilde{x}_{i}\sim U\{\min(\textrm{Income}),\max(\textrm{Income})\}$,
calibrated to Engel's data. The value of $\beta$ is set to the value
of estimates obtained by the method suggested in \citet{Koenker:Xiao:2002}:
for a grid of $R=235$ quantile indices $\{u_{1},\ldots,u_{R}\}$,
$\{\hat{\beta}_{0,QR}(u_{r}),\hat{\beta}_{1,QR}(u_{r})\}^{\textrm{T}}$
are estimated by quantile regression, and $\alpha$ and $\beta$ are
set equal to the estimates obtained from linear regression of $\{\hat{\beta}_{0,QR}(u_{r}),\hat{\beta}_{1,QR}(u_{r})\}^{\textrm{T}}$
on $[\{1,\Phi^{-1}(u_{r})\}:1,\ldots,R]^{\textrm{T}}$, where $\Phi^{-1}$
is the inverse standard normal distribution. We set $\alpha=(86\cdot56,0\cdot55)^{\textrm{T}}$
and $\beta=(-22\cdot17,0\cdot12)^{\textrm{T}}$. Thus the quantile
regression parameters are $\beta_{0}(u)=\alpha_{1}+\alpha_{2}\Phi^{-1}(u)$
and $\beta_{1}(u)=\beta_{1}+\beta_{2}\Phi^{-1}(u)$, and $F_{Y\mid X}(y\mid x)=\Phi\{(y-\alpha_{1}-\beta_{1}\widetilde{x})/(\alpha_{2}+\beta_{2}\widetilde{x})\}$.
As a benchmark, $F_{Y\mid X}(y\mid x)$ is also estimated by applying
the inversion procedure of \citet{CFG:2010} to the quantile regression
process, as $\hat{u}_{i}^{QR}=\epsilon+\int_{\epsilon}^{1-\epsilon}1\{\hat{\beta}_{0,QR}(u)+\hat{\beta}_{1,QR}(u)\widetilde{x}_{i}\leq y_{i}\}du$,
with $\epsilon=0\cdot01$. Dual regression multipliers yield functional
coefficients estimates $\beta_{0}^{*}(u)=(\gamma_{1}^{*}-\lambda_{1}^{*}\bar{x})+(\gamma_{2}^{*}-\lambda_{2}^{*}\bar{x})F_{n}^{-1}(u)$
and $\beta_{1}^{*}(u)=\sum_{j=1}^{J}\lambda_{j}^{*}h_{j}\{F_{n}^{-1}(u)\}$,
where $F_{n}^{-1}$ is the empirical quantile function of $e^{*}$
and $\bar{x}=n^{-1}\sum_{i=1}^{n}\widetilde{x}_{i}$, and with the transformed
intercept coefficients accounting for the centering of $\widetilde{x}_{i}$
in the implementation of (GD).

\begin{table}
\def~{\hphantom{0}}
\caption{$L^{p}$ estimation errors $(\times100)$ and ratios of $L^{p}$
estimation errors $(\times100)$ of $J$ term generalized dual ($L_{GDR(J)}^{p}$)
and quantile regression ($L_{QR}^{p}$)
estimates of $F_{Y\mid X}(y_{i}\mid x_{i})$  $(i=1,\ldots,n)$, for $p=1,2,\infty$
and $J=4,6,8$.}{%
\begin{tabular}{lcccccc}
Sample size & $L_{GDR(4)}^{1}$ & $L_{GDR(4)}^{1}/L_{QR}^{1}$ & $L_{GDR(4)}^{2}$ & $L_{GDR(4)}^{2}/L_{QR}^{2}$ & $L_{GDR(4)}^{\infty}$ & $L_{GDR(4)}^{\infty}/L_{QR}^{\infty}$\\[5pt]
$n=100$  & $4\cdot09$ & $92\cdot89$ & $~5\cdot59$ & $91\cdot87$ & $21\cdot47$ & $88\cdot16$\\
$n=235$  & $2\cdot69$ & $91\cdot34$ & $~3\cdot71$ & $89\cdot19$ & $16\cdot75$ & $80\cdot32$\\
$n=500$  & $1\cdot85$ & $90\cdot42$ & $~2\cdot55$ & $87\cdot52$ & $12\cdot63$ & $73\cdot05$\\
$n=1000$  & $1\cdot31$ & $89\cdot95$ & $~1\cdot82$ & $86\cdot61$ & $~9\cdot68$ & $68\cdot31$\\[15pt]
Sample size & $L_{GDR(6)}^{1}$ & $L_{GDR(6)}^{1}/L_{QR}^{1}$ & $L_{GDR(6)}^{2}$ & $L_{GDR(6)}^{2}/L_{QR}^{2}$ & $L_{GDR(6)}^{\infty}$ & $L_{GDR(6)}^{\infty}/L_{QR}^{\infty}$\\[5pt]
$n=100$  &  $~4\cdot12$ & $93\cdot59$ & $~5\cdot66$ & $93\cdot07$ & $22\cdot30$ & $91\cdot56$\\
$n=235$  &  $ ~2\cdot71$ & $92\cdot03$ & $~3\cdot76$ & $90\cdot43$ & $17\cdot75$ & $85\cdot11$\\
$n=500$  &  $ ~1\cdot86$ & $91\cdot10$ & $~2\cdot59$ & $88\cdot78$ & $13\cdot69$ & $79\cdot19$\\
$n=1000$  &  $ ~1\cdot32$ & $90\cdot64$ & $~1\cdot85$ & $87\cdot90$ & $10\cdot66$ & $75\cdot23$\\[15pt]
Sample size & $L_{GDR(8)}^{1}$ & $L_{GDR(8)}^{1}/L_{QR}^{1}$ & $L_{GDR(8)}^{2}$ & $L_{GDR(8)}^{2}/L_{QR}^{2}$ & $L_{GDR(8)}^{\infty}$ & $L_{GDR(8)}^{\infty}/L_{QR}^{\infty}$\\[5pt]
$n=100$  & $~4\cdot13$ & $93\cdot81$ & $~5\cdot68$ & $93\cdot38$ & $22\cdot44$ & $92\cdot14$\\
$n=235$  &  $~2\cdot71$ & $92\cdot18$ & $~3\cdot77$ & $90\cdot71$ & $17\cdot92$ & $85.\cdot96$\\
$n=500$  &  $~1\cdot86$ & $91\cdot26$ & $~2\cdot60$ & $89\cdot08$ & $13\cdot91$ & $80\cdot46$\\
$n=1000$  &  $~1\cdot32$ & $90\cdot81$ & $~1\cdot86$ & $88\cdot22$ & $10\cdot92$ & $77\cdot01$
\end{tabular}}
\label{tab:3}
\end{table}

\begin{table}
\def~{\hphantom{0}}
\caption{Summary results for intercept and $X$ coefficients across
sample sizes: square root of mean absolute error across simulations (RMAE)
 for $\{0\cdot5,0\cdot9,0\cdot99\}$ quantile indices
and Average (Ave.) RMAE over $\{0\cdot01,0\cdot02,\ldots,0\cdot99\}$
quantile indices.}{%

\begin{tabular}{llccccc}
 & &  \multicolumn{4}{c}{Intercept $\beta_{0}(u)$}\\[5pt]
 Sample size & Method & $\tau=0\cdot5$  & $\tau=0\cdot9$  & $\tau=0\cdot99$ & Ave.\\
 $n=100$ & ~~GDR  & $~8\cdot19$ & $~9\cdot32$ & $10\cdot94$ & $~9\cdot49$\\
 & ~~QR  & $~8\cdot22$ & $~9\cdot38$ & $11\cdot46$ & $~9\cdot69$\\
$n=235$ & ~~GDR  & $~6\cdot59$ & $~7\cdot46$ & $~9\cdot01$ & $~7\cdot69$\\
 & ~~QR  & $~6\cdot63$ & $~7\cdot50$ & $~9\cdot28$ & $~7\cdot81$\\
$n=500$ & ~~GDR  & $~5\cdot47$ & $~6\cdot25$ & $~7\cdot45$ & $~6\cdot39$\\
 & ~~QR  & $~5\cdot50$ & $~6\cdot28$ & $~7\cdot71$ & $~6\cdot50$\\
 $n=1000$  & ~~GDR  & $~4\cdot58$ & $~5\cdot23$ & $~6\cdot33$ & $~5\cdot38$\\
 & ~~QR  & $~4\cdot60$ & $~5\cdot26$ & $~6\cdot50$ & $~5\cdot45$\\[15pt]
&  & \multicolumn{4}{c}{$X$ coefficient $\beta_{1}(u)$}\\[5pt]
 $n=100$ & ~~GDR & $~0\cdot13$ & $~0\cdot15$ & $~0\cdot20$ & $~0\cdot16$\\
 & ~~QR & $~0\cdot14$ & $~0\cdot17$ & $~0\cdot27$ & $~0\cdot19$\\
 $n=235$ & ~~GDR & $~0\cdot11$ & $~0\cdot12$ & $~0\cdot16$ & $~0\cdot13$\\
 & ~~QR & $~0\cdot11$ & $~0\cdot13$ & $~0\cdot20$ & $~0\cdot15$\\
 $n=500$ & ~~GDR & $~0\cdot09$ & $~0\cdot10$ & $~0\cdot14$ & $~0\cdot11$\\
 & ~~QR  & $~0\cdot09$ & $~0\cdot11$ & $~0\cdot17$ & $~0\cdot12$\\
 $n=1000$ & ~~GDR & $~0\cdot07$ & $~0\cdot09$ & $~0\cdot12$ & $~0\cdot09$\\
 & ~~QR & $~0\cdot08$ & $~0\cdot09$ & $~0\cdot14$ & $~0\cdot10$\\[15pt]
\end{tabular}}
\label{tab:4}
\end{table}

Table \ref{tab:BICtab} shows the distribution of selected models
across simulations. The 2 and 4 terms representations are selected
in most simulations, wth the proportion of incorrect selections decreasing
from $25\%$ to $10\%$ as sample size increases. For completeness,
Table \ref{tab:3} reports average estimation errors of conditional
distribution function estimates across simulations for $J=4,6$ and
$8$ terms generalized dual regression and quantile regression-based
estimators, respectively, and their ratio in percentage terms. The
performance of dual regression estimates in the simulations is robust
to incorrect choice of $J$, with only a small loss in accuracy caused
by misspecification. For the 8 terms representation, the gains over
quantile regression-based estimates remain significant, ranging from
$6\%$ to $23\%$ depending on the norm and sample size.

Table \ref{tab:4} summarizes the results corresponding to the accuracy
of functional intercept and covariate coefficients estimates across
simulations. Estimates are based on the selected model in each simulation.
For each coefficient, we compute the root mean absolute error of estimates,
by computing errors for quantile indices in $\{0\cdot5,0\cdot9,0\cdot99\}$
for each replication, and then computing the summary statistic. We
also report average root mean absolute error over the grid $\{0\cdot01,0\cdot02,\ldots,0\cdot99\}$
of quantile indices. In all cases selected generalized dual regression estimates have lower
root mean absolute error, which corroborates results shown in Table
1 in the main text for the conditional distribution function.

\subsection{Additional Simulations\label{sub:Additional-simulations}}

We provide additional simulations comparing dual regression to the
noncrossing quantile regression method introduced by \citet{BRW:2010},
replicating the experiments they propose. In their simulation study
they consider three examples which are special cases of the linear
heteroscedastic model 
\[
y_{i}=\alpha_{1}+\beta_{1}^{\textrm{T}} \widetilde{x}_{i}+(\alpha_{2}+\beta_{2}^{\textrm{T}} \widetilde{x}_{i})\varepsilon_{i},
\]
where each component of $\widetilde{x}_{i}$ satisfies $\widetilde{x}_{ik}\sim U(0,1)$,
$\varepsilon_{i}\sim N(0,1)$, and with $\alpha_{1}=\alpha_{2}=1$.
Their method imposes noncrossing constraints on the quantile regressions
estimated, and they show that it outperforms both linear quantile
regression and the method of \citet{He:1997} in their proposed experiments.
The three examples are:

Example 1. $\textrm{dim}(\widetilde{x}_{i})=4$, $\beta_{1}=(1,1,1,1)^{\textrm{T}}$,
and $\beta_{2}=(0\cdot1,0\cdot1,0\cdot1,0\cdot1)^{\textrm{T}}$.

Example 2. $\textrm{dim}(\widetilde{x}_{i})=10$, $\beta_{1}=(1,1,1,1,0^{\textrm{T}})^{\textrm{T}}$,
and $\beta_{2}=(0\cdot1,0\cdot1,0\cdot1,0\cdot1,0^{\textrm{T}})^{\textrm{T}}$.

Example 3. $\textrm{dim}(\widetilde{x}_{i})=7$, $\beta_{1}=(1,1,1,1,1,1,1)^{\textrm{T}}$,
and $\beta_{2}=(1,1,1,0,0,0,0)^{\textrm{T}}$.

\begin{table}[t]
\def~{\hphantom{0}}
\caption{Replication of \citet{BRW:2010} experiment 1: average root
mean integrated squared error ($\times100$) over 500 simulations,
with standard error in parentheses. NCRQ: noncrossing quantile
regression.
}{%
\begin{tabular}{ccccc}
 &  &  & Example 1  & \\
 &  &  &  & \\
 &  & $\tau=0\cdot5$  & $\tau=0\cdot9$  & $\tau=0\cdot99$\\
 &  &  &  & \\
 &  &  & $n=100$  & \\
GDR  &  & $26\cdot09\,(0\cdot39)$  & $37\cdot04\,(0\cdot54)$  & $57\cdot18\,(0\cdot85)$\\
NCRQ  &  & $29\cdot91\,(0\cdot45)$  & $41\cdot97\,(0\cdot57)$  & $72\cdot24\,(0\cdot88)$\\
GDR (2): Ratio$\times100$  &  & $87\cdot21$ & $88\cdot25$ & $79\cdot16$\\
GDR (4): Ratio$\times100$  &  & $92\cdot56$ & $92\cdot15$ & $80\cdot39$\\
GDR (6): Ratio$\times100$  &  & $93\cdot24$ & $92\cdot98$ & $81\cdot03$\\
GDR (8): Ratio$\times100$  &  & $94\cdot65$ & $93\cdot48$ & $81\cdot29$\\
 &  &  &  & \\
 &  &  & $n=200$  & \\
GDR  &  & $18\cdot80\,(0\cdot27)$  & $25\cdot72\,(0\cdot39)$  & $42\cdot16\,(0\cdot64)$\\
NCRQ  &  & $22\cdot18\,(0\cdot32)$  & $30\cdot03\,(0\cdot46)$  & $57\cdot04\,(0\cdot72)$\\
GDR (2): Ratio$\times100$  &  & $84\cdot76$ & $85\cdot66$ & $73\cdot92$\\
GDR (4): Ratio$\times100$  &  & $91\cdot16$ & $88\cdot31$ & $74\cdot94$\\
GDR (6): Ratio$\times100$  &  & $92\cdot72$ & $89\cdot46$ & $74\cdot96$\\
GDR (8): Ratio$\times100$  &  & $93\cdot27$ & $89\cdot42$ & $75\cdot11$\\
 &  &  &  & \\
 &  &  & $n=500$  & \\
GDR  &  & $12\cdot16\,(0\cdot17)$  & $16\cdot36\,(0\cdot24)$  & $27\cdot21\,(0\cdot43)$\\
NCRQ  &  & $14\cdot32\,(0\cdot20)$  & $19\cdot50\,(0\cdot29)$  & $40\cdot08\,(0\cdot57)$\\
GDR (2): Ratio$\times100$  &  & $84\cdot89$ & $83\cdot89$ & $67\cdot89$\\
GDR (4): Ratio$\times100$  &  & $90\cdot29$ & $86\cdot84$ & $69\cdot03$\\
GDR (6): Ratio$\times100$  &  & $91\cdot60$ & $87\cdot45$ & $69\cdot35$\\
GDR (8): Ratio$\times100$  &  & $92\cdot15$ & $87\cdot96$ & $69\cdot38$
\end{tabular}}
\label{tab:5}
\end{table}

\begin{table}[t]
\def~{\hphantom{0}}
\caption{Replication of \citet{BRW:2010} experiment 2: average root
mean integrated squared error ($\times100$) over 500 simulations,
with standard error in parentheses. NCRQ: noncrossing quantile
regression.
}{%
\begin{tabular}{ccccc}
 &  &  & Example 2  & \\
 &  &  &  & \\
 &  & $\tau=0\cdot5$  & $\tau=0\cdot9$  & $\tau=0\cdot99$\\
 &  &  &  & \\
 &  &  & $n=100$  & \\
GDR &  & $40\cdot24\,(0\cdot40)$  & $56\cdot37\,(0\cdot57)$  & $87\cdot62\,(0\cdot80)$\\
NCRQ  &  & $42\cdot55\,(0\cdot43)$  & $53\cdot18\,(0\cdot49)$  & $90\cdot30\,(0\cdot84)$\\
GDR (2): Ratio$\times100$  &  & $94\cdot58$ & $106\cdot00$ & $97\cdot03$\\
GDR (4): Ratio$\times100$  &  & $105\cdot02$ & $110\cdot83$ & $100\cdot26$\\
GDR (6): Ratio$\times100$  &  & $110\cdot37$ & $113\cdot69$ & $101\cdot87$\\
GDR (8): Ratio$\times100$  &  & $124\cdot84$ & $123\cdot50$ & $105\cdot45$\\
 &  &  &  & \\
 &  &  & $n=200$  & \\
GDR  &  & $28\cdot63\,(0\cdot28)$  & $39\cdot03\,(0\cdot37)$  & $60\cdot34\,(0\cdot61)$\\
NCRQ  &  & $31\cdot48\,(0\cdot31)$  & $39\cdot99\,(0\cdot38)$  & $66\cdot98\,(0\cdot63)$\\
GDR (2): Ratio$\times100$  &  & $90\cdot93$ & $97\cdot59$ & $90\cdot10$\\
GDR (4): Ratio$\times100$  &  & $98\cdot25$ & $102\cdot37$ & $91\cdot20$\\
GDR (6): Ratio$\times100$  &  & $100\cdot57$ & $103\cdot43$ & $91\cdot56$\\
GDR (8): Ratio$\times100$  &  & $102\cdot00$ & $104\cdot04$ & $91\cdot73$\\
 &  &  &  & \\
 &  &  & $n=500$  & \\
GDR  &  & $17\cdot78\,(0\cdot17)$  & $24\cdot23\,(0\cdot23)$  & $37\cdot02\,(0\cdot39)$\\
NCRQ  &  & $20\cdot87\,(0\cdot20)$  & $27\cdot86\,(0\cdot26)$  & $47\cdot65\,(0\cdot43)$\\
GDR (2): Ratio$\times100$  &  & $85\cdot19$ & $86\cdot98$ & $77\cdot69$\\
GDR (4): Ratio$\times100$  &  & $91\cdot82$ & $90\cdot59$ & $79\cdot60$\\
GDR (6): Ratio$\times100$  &  & $93\cdot49$ & $91\cdot74$ & $79\cdot99$\\
GDR (8): Ratio$\times100$  &  & $94\cdot33$ & $92\cdot28$ & $80\cdot07$
\end{tabular}}
\label{tab:6}
\end{table}

For each example, 500 datasets of size 100, 200 and
500 are simulated. For the method of \citet{BRW:2010}, six quantile
curves are fitted to the data for each example, $u=\{0\cdot1,0\cdot3,0\cdot5,0\cdot7,0\cdot9,0\cdot99\}$.
We also implemented the noncrossing quantile regression method by
fitting eleven quantile curves for the larger sequence $u=\{0\cdot01,0\cdot1,0\cdot2,\ldots,0\cdot9,0\cdot99\}$,
the results are similar and are thus omitted.

Tables \ref{tab:5}--\ref{tab:7} show the average root mean integrated
squared errors over the 500 datasets along with their estimated standard
errors, for each sample size, and for each of $u=\{0\cdot5,0\cdot9,0\cdot99\}$.
For each simulation, the empirical root mean integrated squared error
is calculated as $\textrm{RMISE}=[n^{-1}\sum_{i=1}^{n}\{\hat{\beta}(u)^{\textrm{T}} x_{i}-\beta(u)^{\textrm{T}} x_{i}\}^{2}]^{1/2}$,
where $\hat{\beta}(u)$ and $\beta(u)$ are the estimated and true
vector of quantile regression coefficients, respectively. The results
for the other quantiles are similar, and are thus omitted.

In all three examples the location-scale structure, $J=2$, is selected
by the Schwartz criterion for each simulation and our proposed estimator
significantly outperforms the noncrossing quantiles method for all
quantiles and all sample sizes, except for $n=100$ and $\tau=0\cdot9$
in Example 2. The good relative performance of dual regression results
from the selected location-scale structure, which adds further smoothness
and stability across quantile curves, beyond the noncrossing constraints
imposed by noncrossing quantile regression. This improvement is greater
in the tails, as dual regression solutions are estimated globally
whereas the local nature of quantile regression affects estimation
of extreme quantiles. 

We also report the relative performance of non-selected dual regression
estimates. Apart from Examples 2 and 3 with $n=100$, the results
are similar for all $J$ to the selected model $J=2$. For $n=100$,
results for Example 2, and to a lesser extent Example 3, show that
the relative performance of dual regression deteriorates, especially
for $J=8$. These results are driven by a few simulations where the
solver was unable to find an optimal solution, 7 instances for Example
2 and 5 for Example 3. Since for Example 2 and $J=8$ the number
of parameters is $2+8\times10=82$ for 100 observations, this is not
unexpected. Compared to the simulations calibrated to the Engel data
example, the fact that representations with $J$ greater 2 are never
selected for Examples 1--3 suggest that the presence of multiple covariates
provides useful information effectively accounted for by the proposed
model selection procedure.

%\clearpage

\begin{table}[t]
\def~{\hphantom{0}}
\caption{Replication of \citet{BRW:2010} experiment 3: average root
mean integrated squared error ($\times100$) over 500 simulations,
with standard error in parentheses. NCRQ: noncrossing quantile
regression.
}{%
\begin{tabular}{ccccc}
 &  &  & Example 3  & \\
 &  &  &  & \\
 &  & $\tau=0\cdot5$  & $\tau=0\cdot9$  & $\tau=0\cdot99$\\
 &  &  &  & \\
 &  &  & $n=100$  & \\
GDR  &  & $69\cdot04\,(0\cdot84)$  & $95\cdot85\,(1\cdot14)$  & $152\cdot77\,(1\cdot75)$\\
NCRQ  &  & $75\cdot09\,(0\cdot85)$  & $97\cdot98\,(1\cdot20)$  & $178\cdot10\,(2\cdot01)$\\
GDR (2): Ratio$\times100$  &  & $~91\cdot94$ & $~97\cdot82$ & $~85\cdot78$\\
GDR (4): Ratio$\times100$  &  & $99\cdot80$ & $102\cdot94$ & $~87\cdot23$\\
GDR (6): Ratio$\times100$  &  & $102\cdot96$ & $104\cdot99$ & $~88\cdot09$\\
GDR (8): Ratio$\times100$  &  & $105\cdot49$ & $106\cdot96$ & $~88\cdot29$\\
 &  &  &  & \\
 &  &  & $n=200$  & \\
GDR  &  & $49\cdot22\,(0\cdot56)$  & $67\cdot26\,(0\cdot77)$  & $105\cdot22\,(1\cdot30)$\\
NCRQ  &  & $55\cdot12\,(0\cdot62)$  & $72\cdot82\,(0\cdot83)$  & $135\cdot24\,(1\cdot59)$\\
GDR (2): Ratio$\times100$  &  & $~89\cdot29$ & $~92\cdot37$ & $~77\cdot80$\\
GDR (4): Ratio$\times100$  &  & $~95\cdot78$ & $~95\cdot74$ & $~79\cdot65$\\
GDR (6): Ratio$\times100$  &  & $~97\cdot41$ & $~97\cdot06$ & $~80\cdot22$\\
GDR (8): Ratio$\times100$  &  & $~98\cdot51$ & $~97\cdot08$ & $~80\cdot24$\\
 &  &  &  & \\
 &  &  & $n=500$  & \\
GDR  &  & $30\cdot84\,(0\cdot35)$  & $42\cdot17\,(0\cdot51)$  & $66\cdot86\,(0\cdot89)$\\
NCRQ  &  & $35\cdot77\,(0\cdot42)$  & $48\cdot80\,(0\cdot57)$  & $94\cdot12\,(1\cdot19)$\\
GDR (2): Ratio$\times100$  &  & $~86\cdot22$ & $~86\cdot40$ & $~71\cdot04$\\
GDR (4): Ratio$\times100$  &  & $~92\cdot60$ & $~90\cdot27$ & $~72\cdot66$\\
GDR (6): Ratio$\times100$  &  & $~94\cdot38$ & $~91\cdot20$ & $~72\cdot73$\\
GDR (8): Ratio$\times100$  &  & $~94\cdot85$ & $~91\cdot27$ & $~72\cdot82$
\end{tabular}}
\label{tab:7}
\end{table}

\end{document}